\documentclass[12pt]{article}

\usepackage{latexsym}
\usepackage{amssymb,amsfonts,amsmath}
\usepackage{graphicx}
\usepackage{indentfirst}
\usepackage{bbm}
\usepackage{enumitem}
\usepackage{amssymb}
\usepackage{verbatim}
\usepackage{amsmath, amsthm,amssymb}
\usepackage{mathrsfs}
\usepackage{hyperref}
\usepackage{amsfonts}
\usepackage{dsfont}
\usepackage{cite}
\usepackage{xcolor}
\usepackage[open,openlevel=2,atend,numbered]{bookmark}
\usepackage[multiple]{footmisc}
\usepackage[makeroom]{cancel}
\usepackage[normalem]{ulem}
\usepackage{stackengine}
\stackMath
\usepackage{tikz}
\usepackage{braket}

\parskip=\medskipamount

\counterwithin*{equation}{section}

\newcommand{\be}{\begin{equation}}
\newcommand{\ee}{\end{equation}}
\newcommand{\bea}{\begin{eqnarray}}
\newcommand{\eea}{\end{eqnarray}}
\def\baa#1\eaa{\begin{align}#1\end{align}}
\def\bna#1\ena{\begin{align*}#1\end{align*}}
\def\ben#1\een{\begin{enumerate}#1\end{enumerate}}
\newcommand{\non}{\nonumber}
\newcommand {\cG}{{\cal G}}

\newcommand {\cI}{{\cal I}}

\newcommand {\cN}{{\cal N}}
\newcommand {\cO}{{\cal O}}

\renewcommand{\a}{\alpha}
\renewcommand{\b}{\beta}
\renewcommand{\c}{\chi}
\renewcommand{\d}{\delta}

\newcommand{\g}{\gamma}
\newcommand{\h}{\eta}

\renewcommand{\k}{\kappa}
\renewcommand{\l}{\lambda}

\newcommand{\p}{\pi}

\renewcommand{\r}{\rho}

\renewcommand{\t}{\tau}

\newcommand{\D}{\Delta}

\newcommand{\G}{\Gamma}
\renewcommand{\L}{\Lambda}

\newcommand{\Tr}{\text{Tr}}
\newcommand{\tr}{\text{tr}}
\newcommand{\arctanh}{\text{arctanh}}

\newcommand{\rd}{\text{d}}
\newcommand{\ri}{\text{i}}
\newcommand{\re}{\text{e}}

\newcommand{\dsR}{{\mathbb R}}
\newcommand{\dsC}{{\mathbb C}}
\newcommand{\dsQ}{{\mathbb Q}}

\newcommand{\dsG}{{\mathbb G}}
\newcommand{\dsP}{{\mathbb P}}
\newcommand{\dsV}{{\mathbb V}}
\newcommand{\dsT}{{\mathbb T}}
\newcommand{\dsH}{{\mathbb H}}
\newcommand{\dsM}{{\mathbb M}}
\newcommand{\dsA}{{\mathbb A}}
\newcommand{\dsB}{{\mathbb B}}

\newcommand{\dsO}{{\mathbb O}}

\newcommand{\sU}{\mathsf{U}}

\newcommand{\ph}{\phantom}
\newcommand{\id}{\mathds{1}}
\renewcommand{\non}{\nonumber}

\newcommand{\pd}{\partial}

\newcommand{\blue}{\textcolor{blue}}

\newcommand{\inte}{\int\!\!\rd}

\newcommand{\ve}{\varepsilon}

\newcommand{\hf}{\frac12}

\begin{document}

\begin{titlepage}
\begin{flushright}
January, 2026\\
\end{flushright}
\vspace{5mm}

\begin{center}
{\Large \bf Heat kernel approach to the one-loop effective action for nonlinear electrodynamics}
\end{center}

\begin{center}

{\bf Evgeny I. Buchbinder, Darren T. Grasso and Joshua R. Pinelli}
\vspace{5mm}

\footnotesize{
	{\it Department of Physics M013, The University of Western Australia\\
		35 Stirling Highway, Crawley W.A. 6009, Australia}}
\vspace{2mm}
~\\
Email: \texttt{evgeny.buchbinder@uwa.edu.au, darren.grasso@uwa.edu.au,\\ joshua.pinelli@research.uwa.edu.au}\\
\vspace{2mm}

\end{center}

\begin{abstract}
{\baselineskip=14pt
	
We develop a heat kernel method to compute the one-loop effective action for a general class of nonlinear electrodynamic (NLED) theories in four dimensional Minkowski spacetime. 
Working in the background field formalism, we extract the logarithmically divergent part of the effective action, the so-called \textit{induced action}, corresponding to the 
DeWitt $a_2$ coefficient of the heat kernel. In NLED, quantisation yields non-minimal differential operators, for which standard heat kernel techniques are not immediately applicable. 
Considering the weak-field regime, we calculate the $a_0$, $a_1$ and $a_2$ contributions to leading order in the background electromagnetic field strength. 
Finally, we consider conformal NLED theories and compute the $a_0$ contribution to all orders.
For this class, we comment on the role of causality being necessary and sufficient for the convergence of the exact $a_1$ and $a_2$ contributions. }
\end{abstract}

\vfill

\vfill
\end{titlepage}

\newpage
\renewcommand{\thefootnote}{\arabic{footnote}}
\setcounter{footnote}{0}

\tableofcontents{}
\vspace{1cm}
\bigskip\hrule

\allowdisplaybreaks


\section{Introduction}

The study of gauge invariant nonlinear electrodynamics (NLED) began with the seminal work by Born and Infeld (1934) \cite{Born:1934fnft}, motivated by a desire to resolve the infinite self-energy of point charges that arise in Maxwell’s theory. Soon after, the possibility of nonlinear photon-photon interactions was investigated in quantum electrodynamics and the one-loop effective action was derived to all orders in a constant electromagnetic background~\cite{Euler:1935zz, Heisenberg:1936nmg}. In the 80s Born--Infeld theory re-emerged in the context of string theory. For bosonic open strings, Fradkin and Tseytlin (1985) \cite{Fradkin:1985qd} showed that the Born--Infeld action arises as a low-energy effective action \cite{Tseytlin:1986ti, Abouelsaood:1986gd, Bergshoeff:1987at, Metsaev:1987qp, Dai:1989ua}. In both bosonic and supersymmetric settings, open and type II string effective actions generate explicit higher-derivative corrections to Born--Infeld theory. These include two derivative \cite{Andreev:1988cb, Hashimoto:1999jx, Andreev:2001bo} and four derivative terms \cite{Andreev:1988cb, Leigh:1989jq, Shmakova:1999ai, Wyllard:2000qe, BabaeiVelni:2019ptj}. Equivalent four derivative and quartic field strength structures resurface in other supersymmetric contexts 
(see e.g. \cite{DeGiovanni:1999hr, Tseytlin:1999dj, Koerber:2002zb, Chemissany:2006qd, Grasso:2007zfa}). 

NLED in four dimensions is formulated in terms of a Lagrangian \cite{Landau:1951ll,Peres:1961zz,Plebanski:1968nl,Plebanski:1970zz,Boillat:1970gw} which is a scalar function of the two electromagnetic Lorentz invariants, first introduced by Minkowski \cite{Minkowski:1908ed}:
\be \label{NLEDLagrangian}
	L_{\rm NLED}(F)=L(\a,\b)~,\qquad \a:=\tfrac{1}{4}F^{ab}F_{ab}~,\qquad \b:=\tfrac{1}{4}\~F^{ab}F_{ab}~.
\ee
Here $\~F^{ab}=\tfrac{1}{2} \ve^{abcd}F_{cd}$ is the Hodge dual of the electromagnetic field strength $F_{ab}=2\pd_{[a}A_{b]}$, with $\ve^{abcd}$ the Levi--Civita symbol. For example, the Born--Infeld Lagrangian takes the form
\be\label{eqn:BornInfeldLagrangian}
	L_{\rm BI}(\a,\b)=T-T\sqrt{-\det\big(\h_{ab}+T^{-\hf}F_{ab}\big)}=T-\sqrt{T^2+2\a T-\b^2}~,
\ee
where $T$ is a coupling constant with dimensions of energy density. In general, we may assume zero vacuum energy and a non-vanishing kinetic term:
\be
	L(0,0)=0~,\qquad\qquad L_{\a}:=\frac{\pd L(\a,\b)}{\pd \a} \neq 0~.
\ee
The latter ensures that the linearised field equations propagate two independent polarisation modes. For a broader introduction to nonlinear electrodynamics and its various formulations,  see e.g.~\cite{Sorokin:2021tge}
for a  review.

Deeper insights into the structure of NLED came from attempting to retain the symmetries inherent to Maxwell electrodynamics. These are duality invariance of the equations of motion \cite{Rainich:1925qbz} and conformal invariance of the Lagrangian \cite{Bateman:1909pyp, Bateman:1910mvi, Cunningham:1910pxu}. For a generic NLED Lagrangian~\eqref{NLEDLagrangian} the condition of $\sU(1)$ duality invariance can be written as \cite{Bialynicki-Birula:1984daz, Gibbons:1995cv, Gibbons:1995ap, 
	Gaillard:1997zr, Gaillard:1999gz}
\be 
	\b(L_{\b}^2-L_{\a}^2-1)=-2\a L_{\a}L_{\b}~.
	\label{eqn:DualityInvariance}
\ee
Here and throughout this work we use  the following notation for the first and second derivatives of the Lagrangian:
\baa\label{aln:FirstandSecondDerivsNotation}
	L_{\a}&:=\tfrac{\pd L(\a,\b)}{\pd\a}~,\quad L_{\b}:=\tfrac{\pd L(\a,\b)}{\pd\b}~,\\
	L_{\a\a}&:=\tfrac{\pd^2 L(\a,\b)}{\pd\a^2}~,\quad L_{\a\b}:=\tfrac{\pd^2 L(\a,\b)}{\pd\a\,\pd\b}~, \quad L_{\b\b}:=\tfrac{\pd^2 L(\a,\b)}{\pd\b^2}~.\label{aln:FirstandSecondDerivsNotation2}
\eaa
Following~\cite{Gaillard:1997zr, Gaillard:1999gz,Kuzenko:2000tg,Kuzenko:2000uh}, we refer to NLED theories satisfying $\sU(1)$ duality invariance as \emph{self-dual}. An example of a self-dual theory is Born--Infeld electrodynamics \eqref{eqn:BornInfeldLagrangian}, but it is not conformally invariant \cite{Schrodinger:1935oqa}. Stemming from a symmetric, traceless energy-momentum tensor is the condition for conformal invariance of an NLED theory \cite{Kosyakov:2007qc, Denisov:2017qou, Bandos:2021rqy}
\be \label{eqn:ConformalNLEDProperty}
	\a L_{\a}+\b L_{\b}=L \qquad \implies \qquad L_{\a\a}L_{\b\b}-L_{\a\b}^2=0~.
\ee
We will refer to any NLED theory satisfying this condition as \emph{a conformal NLED theory}. With the exception of Maxwell electrodynamics, every conformal NLED theory is non-analytic. The unique one parameter deformation of Maxwell electrodynamics which preserves both conformal and duality invariance was recently constructed, ModMax electrodynamics \cite{Bandos:2020jsw} (also see \cite{Kosyakov:2020wxv}). It is described by the Lagrangian
\be
	L_{\rm MM}(\a,\b)=- \a\cosh \g  + \sqrt{\a^2+\b^2}\,\sinh \g ~,
\ee
with $\g\geq0$ a dimensionless coupling constant.

Of significant interest in the context of NLED is the notion of causality. Although in their early work Pleba\'nsky \cite{Plebanski:1968nl, Plebanski:1970zz} and Boillat \cite{Boillat:1970gw} identified sufficient conditions for causal wave propagation and the absence of birefringence, the complete set of necessary and sufficient conditions was derived only recently in~\cite{Schellstede:2016zue}. Their results guarantee that classical wavefronts propagate subluminally on nonlinear electromagnetic backgrounds, preserving hyperbolicity and consistency with special relativity. In~\cite{Russo:2024kto, Russo:2024llm, Russo:2024xnh} these results were extended to a curved space background, where causality mirrors both the dominant and strong energy condition on the energy-momentum tensor.

Following~\cite{Russo:2024kto, Russo:2024llm, Russo:2024xnh}, we will distinguish the two types of causality conditions in \cite{Schellstede:2016zue}, assuming the standard sign convention
\be \label{eqn:SignofFirstDeriv}
	L_{\a}<0~.
\ee
The first type consists of the weak-field causality conditions
\be \label{eqn:IntroConvexCondition}
	L_{\a\a}\geq0~,\qquad L_{\b\b}\geq0~,\qquad L_{\a\a}L_{\b\b}-L_{\a\b}^2\geq 0~,
\ee
which are equivalent to the convexity of the Lagrangian with respect to the electric field~\cite{Bandos:2021rqy}.\footnote{For this reason, they may be referred to as the convexity conditions in the literature.} 
The second type is the strong-field causality condition
\be \label{eqn:IntroCausalCondition}
	-L_{\a}+\a(L_{\b\b}-L_{\a\a})-2\b L_{\a\b}-(L_{\a\a}+L_{\b\b})\sqrt{\a^2+\b^2}>0~.
\ee
We will refer to any NLED theory satisfying both causality constraints as \emph{a causal NLED theory}. For self-dual theories with a weak-field limit, the weak-field conditions imply the strong-field condition~\cite{Russo:2024xnh}.
An interesting corollary \cite{Schellstede:2016zue} of the strong-field causality condition occurs for NLED theories independent of $\b$, i.e. $L(\a)$. In such a case (\ref{eqn:IntroCausalCondition}) reduces to
\be
	-L_{\a}-\a L_{\a\a}>L_{\a\a}\sqrt{\a^2+\b^2}~.
\ee
This inequality must fail for sufficiently large $\b$, implying that any $L(\a)$ theory is necessarily acausal, 
except for Maxwell's linear theory (since $L_{\a\a}=0$). Similar arguments~\cite{Russo:2024kto} show that theories of the form $L(\a,\b)=f(\a)+g(\b)$ are likewise acausal.

This paper is devoted to developing a general approach to the quantisation of NLED. 
The quantisation of NLED remains an outstanding problem due to the presence of non-minimal operators. One may classify general matrix valued second-order differential operators into two distinct classes: (i) minimal operators of the form\footnote{We use double-struck Latin letters (e.g. $\dsA$, $\dsB$, $\dsC$) to denote quantities carrying additional matrix indices.}
	\be
		\D=\id\Box+\dsV^{a}\pd_a+\dsT~,\qquad\qquad\quad~~ \Box:=\pd^a\pd_a~,
\label{e1}		
	\ee
for which the Schwinger--DeWitt formalism \cite{Schwinger:1951givp, DeWitt:1964dtgf, Barvinsky:1985gsdt, Buchbinder:1998imss, Avramidi:2000hkqg, DeWitt:2003gaqft, Vassilevich:2003hke} 
yields standard, well-known results; and (ii) non-minimal operators
	\be
		\D=\dsH^{ab}\pd_a\pd_b+\dsV^{a}\pd_a+\dsT~,
		\label{e2}
	\ee
where $\dsH^{ab}$ is a field-dependent matrix  of general form.

Some progress exists for select choices of operators with a non-minimal principle term
$\dsH^{ab}$ \cite{Barvinsky:1985gsdt, Gusynin:1989ky, Gusynin:1991mk, Gusynin:1997dc, Moss:2013cba, Iochum:2016ynh, Iochum:2017ver, Barvinsky:2021ijq, Grasso:2023qye, Barvinsky:2025jbw, Sauro:2025sbt}, 
but no general solution is known. In many cases, a non-minimal operator can be equivalently realised as a minimal operator by a suitable choice of gauge fixing and/or a field 
redefinition within the path integral (see e.g. \cite{Grasso:2023hmv}). However, these methods fail for the non-minimal operator that arises in the quantisation of NLED, which we will later show is of the form:
\be
	\D^{ab}_{\rm NLED}:=-L_{\a}\h^{ab}\Box+G^{acdb}\pd_c\pd_d + V^{acb}\pd_c~,
	\label{e3}
\ee
with $\h^{ab}$ the Minkowski metric.  The main problem is the presence of the tensor $G^{abcd}$, which contains second derivatives of the Lagrangian:
\be 
	G^{abcd}:=\big(L_{\a\a}F^{ab}+L_{\a\b}\~F^{ab}\big)F^{cd}+\big(L_{\a\b}F^{ab}+L_{\b\b}\~F^{ab}\big)\~F^{cd}~.\label{eqn:GTensorDefn}
\ee
The aim of this paper is to develop heat kernel techniques for computing the DeWitt coefficients of this non-minimal operator. Our procedure is based on the Volterra-series approach used previously in 
Refs.~\cite{Iochum:2016ynh, Iochum:2017ver, Grasso:2023qye} to split the exponential of operators which appear in the heat kernel expansion.
Working in the weak-field regime, we compute the leading contributions to the heat kernel, by expanding up to quartic order in the background electromagnetic field strength.
In the case of conformal NLED, we present arguments that the strong-field causality condition~\eqref{eqn:IntroCausalCondition} is necessary for 
convergence of the heat kernel expansion.
We also show, in this case, that satisfying both the weak and strong-field causality conditions is sufficient for convergence.

The paper is organised as follows. In Section \ref{sct:QuantisationNLED} we derive the NLED operator governing the one-loop effective action, obtained by integrating out the quantum vector fields. 
We show it is a non-minimal operator of the general form \eqref{e3}, and all previously developed quantisation schemes are not directly applicable. 
In Section \ref{sct:QuarticHeatKernelTechniques} we develop techniques for computing the heat kernel coefficients associated with NLED. To illustrate the method we compute the heat kernel coefficients to leading order in the electromagnetic field. 
In Section \ref{sct:ConformalHeatKernelTechniques} we consider conformal NLED theories and compute the $a_0$ contribution to all orders. For this case, we also discuss convergence of the Volterra series and demonstrate that it is directly related to the causality conditions.
Generalisations and open problems are briefly discussed in Section \ref{sct:Outlook}. The main body of this paper is accompanied by three technical appendices. 
Appendix \ref{app:4x4MatrixProperties} presents notable matrix properties specific to four dimensions. Appendix \ref{app:DerivativeBasis} constructs a basis for four derivative, 
quartic field strength structures which arise for the induced action of Born--Infeld theory. Finally, Appendix \ref{app:PropertiesofTensorG} compiles properties of 
the tensor $G_{abcd}$ (\ref{eqn:GTensorDefn}), relevant to Sections \ref{sct:QuarticHeatKernelTechniques} and \ref{sct:ConformalHeatKernelTechniques}.


\section{Quantisation of nonlinear electrodynamics}\label{sct:QuantisationNLED}

In this section we identify the one-loop contributions to the background effective action, $\G [A_B]$, defined by integrating out the quantum vector field
\be \label{eqn:PathIntegralFormulation}
	\re^{\ri \G [A_B]} =\int [{\mathfrak D} A_Q] \,
	\d \Big( \varphi - \c (A_Q) \Big) \, {\rm Det} \,(\D_{\rm gh})\,
	\re^{\ri S[A_B+A_Q] }~.
\ee
Here $A_B$ and $A_Q$ are the background and quantum vector fields, respectively, $S[A]$ is the classical action in Minkowski space, corresponding to (\ref{NLEDLagrangian})
\be \label{eqn:NLEDAction}
	S[A]=  \inte^4x\,L(\a,\b)~,
\ee
$\c(A_Q)$ denotes a gauge fixing condition, $\D_{\rm gh}$ the corresponding Faddeev-Popov operator \cite{Fadeev:1967fdym}, and $\varphi$ an arbitrary background scalar field.

In general, the background one-loop effective action in Minkowski space can be obtained from a given theory as the functional trace of some differential operator $\D$
\be \label{eqn:1LoopEffectiveAction}
	\G^{(1)}[A_B]=\frac{\ri}{2}\Tr\ln\D~.
\ee

\subsection{Background-quantum splitting}
By splitting the vector field $A_a$ into background (subscript $B$) and quantum components (subscript $Q$) \cite{Barvinsky:2015hkbff}
\be 
	A_a=A_{B,a}+A_{Q,a}~,
\ee
the classical action (\ref{eqn:NLEDAction}) is expanded up to the one-loop approximation as
\be
	S[A_B+A_Q]=S[A_B]+S_2[A_B,A_Q]~,
\ee
where $S[A_B]$ is the classical action evaluated on the background fields and $S_2[A_B,A_Q]$ represents the term quadratic in quantum fields.\footnote{The term linear in quantum fields vanishes, since the background vector fields are assumed on-shell.} 

If the classical Lagrangian $L(\a,\b)$ (\ref{NLEDLagrangian}) describes a causal NLED model with the causality constraints (\ref{eqn:SignofFirstDeriv}), (\ref{eqn:IntroConvexCondition}) and (\ref{eqn:IntroCausalCondition}), the background Lagrangian $L(\a_B,\b_B)$ associated with $S[A_B]$ inherits these constraints. In terms of the background fields, these constraints are: the background sign convention\footnote{Here we have introduced notation for the first and second derivatives of the background Lagrangian in analogy to the notation introduced in (\ref{aln:FirstandSecondDerivsNotation}) and \eqref{aln:FirstandSecondDerivsNotation2}.}
\be\label{eqn:BackgroundSignConvention}
L_{\a_B}<0~;
\ee
the weak-field background  causality conditions
\be \label{eqn:BackgroundConvexCondition}
L_{\a_B\a_B}\geq0,\qquad L_{\b_B\b_B}\geq0,\qquad L_{\a_B\a_B}L_{\b_B\b_B}-L_{\a_B\b_B}^2\geq 0~;
\ee
and the strong-field causality condition
\be \label{eqn:BackgroundCausalCondition}
-L_{\a_B}+\a_B(L_{\b_B\b_B}-L_{\a_B\a_B})-2\b L_{\a_B\b_B}-(L_{\a_B\a_B}+L_{\b_B\b_B})\sqrt{\a_B^2+\b_B^2}>0~.
\ee

The quadratic action $S_2[A_B,A_Q]$ has a corresponding Lagrangian of the form
\baa\label{aln:QuadraticQuantumLagrangian}
	L_2[A_B,A_Q]&=L_{\a_B}\a_Q+L_{\b_B}\b_Q+\frac{1}{8}\,L_{\a_B\a_B}(F_B^{ab}F_{Q,ab})^2\non\\	&\qquad+\frac{1}{4}\,L_{\a_B\b_B}(F_B^{ab}F_{Q,ab})\big(\~F_B^{cd}F_{Q,cd}\big)+\frac{1}{8}\,L_{\b_B\b_B}\big(\~F_B^{ab}F_{Q,ab}\big)^2~.
\eaa
The action $S_2[A_B,A_Q]$ corresponding to (\ref{aln:QuadraticQuantumLagrangian}) may be recast  into the following form
\be
	S_2[A_B,A_Q]=\hf\inte^4x\,A_{Q,a}\D_2^{ab}A_{Q,b}~,
\ee
from which the quadratic operator $\D_2^{ab}$ can be determined. After integrating (\ref{aln:QuadraticQuantumLagrangian}) by parts, the operator is found to be non-minimal
\baa \label{eqn:VectorOperatorNoGauging}
	\D_2^{ab}&= -L_{\a_B}\h^{ab}\Box+L_{\a_B}\pd^a\pd^b+G_B^{acdb}\pd_c \pd_d\non\\
	&+\Bigl[-(\pd^cL_{\a_B})\h^{ab}+(\pd^bL_{\a_B})\h^{ac}-(\pd_dL_{\b_B})\ve^{abcd}+(\pd_d G_B^{adcb})\Bigr]\pd_c~,
\eaa
with $\Box:=\pd^a\pd_a$ the d'Alembertian, and $G_B^{abcd}$ a background field tensor which contains second derivatives of the Lagrangian:
\be \label{eqn:BackgroundGTensorDefn}
	G_B^{abcd}:=\big(L_{\a_B\a_B}F_B^{ab}+L_{\a_B\b_B}\~F_B^{ab}\big)F_B^{cd}+\big(L_{\a_B\b_B}F_B^{ab}+L_{\b_B\b_B}\~F_B^{ab}\big)\~F_B^{cd}~.
\ee
In Appendix \ref{app:PropertiesofTensorG} we explore key properties of the above background tensor which will prove useful later. 
\subsection{Gauge fixing}

In order for the non-minimal operator (\ref{eqn:VectorOperatorNoGauging}) to be realised as minimal, both the second and third terms would need to be eliminated by an appropriate gauge-fixing condition. No such condition exists, however, and so we restrict ourselves to eliminating the second term by choosing
\be
	\c(A_Q)=\pd^aA_{Q,a}~,
\ee
which leads to a minimal ghost operator that does not contribute to the effective action
\be
	\D_{\rm gh}=\Box~.\label{eqn:GhostOperator}
\ee
Since the effective action (\ref{eqn:PathIntegralFormulation}) is independent of $\varphi$, this field can be integrated out with some weight, which is chosen to be
\be
	{\rm exp} \Bigg(\frac{\ri}{2} \inte^4x\, L_{\a_B}\varphi^2 \Bigg)~.
	\label{weight}
\ee
This leads to the gauge fixing term
\be
	S_{\rm gf}[A_B,A_Q]=\hf \inte^4x\, \,A_{Q,a}\D_{\rm gf}^{ab}A_{Q,b}~,
\ee
where the gauge fixing operator is found to be
\be
	\D_{\rm gf}^{ab}=-L_{\a_B}\pd^a\pd^b-(\pd^aL_{\a_B})\pd^b~.
\ee
The one-loop effective action (\ref{eqn:1LoopEffectiveAction}) is then specified by
\be
	\G^{(1)}[A_B]=\frac{\ri}{2} \Tr\ln{\D}~,
\ee
where the total vector field operator is defined by
\baa
	\D^{ab}:&=\D^{ab}_2+\D^{ab}_{\rm gf}\non\\
	&=-L_{\a_B}\h^{ab}\Box+G_B^{acdb}\pd_c\pd_d + V_B^{acb}\pd_c~, \label{eqn:Non-MinOperatorDefn}
\eaa
having introduced the following notation for the background term linear in spacetime derivatives
\be \label{eqn:LinearTermDef}
	V_B^{acb}:=-(\pd^cL_{\a_B})\h^{ab}+(\pd^bL_{\a_B})\h^{ac}-(\pd^aL_{\a_B})\h^{bc}+(\pd_dL_{\b_B})\ve^{adcb}+(\pd_dG_B^{adcb})~.
\ee


\section{Heat kernel approach to nonlinear electrodynamics}\label{sct:QuarticHeatKernelTechniques}

In this section we study the heat kernel $K(x,x';s)$ for nonlinear electrodynamics, related to the one-loop effective action by
\be \label{eqn:HeatKernelEARelash}
	\G^{(1)}[A_B]=-\frac{\ri}{2}\int^\infty_0\frac{\rd s}{s}\inte^4x\,K(x;s)~,\qquad K(x;s):=\lim_{x'\rightarrow x} \tr\, K(x,x';s)~.
\ee

The heat kernel in the coincidence limit $K(x;s)$ admits the well-known asymptotic expansion in proper time $s$
\be \label{eqn:DeWittCoefficientsFlat}
	K(x;s)=\frac{h}{s^2}\sum_{n=0}^{\infty}(\ri s)^n \, a_n(x)~, \qquad h:=\frac{\ri}{(4\p \ri)^2}~,
\ee
where $a_n(x)$ denotes the matrix trace of the $n^{\rm th}$ DeWitt coefficient in the coincidence limit. For a massless theory, the logarithmically divergent part of the effective action has the form
\be
	\G_{\rm div}[A_B]=-\frac{\ln \L}{(4\p)^2}\inte^4x\,a_2(x)~.
\ee
We identify the induced action with $\inte^4x\,a_2(x)$, modulo an overall numerical coefficient.

Traditional approaches with Fourier methods compute the $a_n$ coefficients using Baker--Campbell--Hausdorff expansions and Gaussian integrations (e.g. \cite{McArthur:1997gaea,Kuzenko:2003bfmop}). However, these techniques break down for the non-minimal operator that arises in NLED quantisation. Instead, we employ a Volterra-series method \cite{Iochum:2016ynh,Iochum:2017ver,Grasso:2023qye}, which has recently proven effective for the non-minimal operator of the form
\be 
	\D_{\rm NM}=\dsM\Box+\dsV^{a}\pd_a+\dsT~,
\ee
for $\dsM,\,\dsV^{a},\,\dsT$ arbitrary matrices. The central idea is to expand the exponential of $s$-dependent operators using the Volterra identity
\baa 
	\re^{\dsA+\dsB(s)}=\re^\dsA+\re^\dsA\sum_{n=1}^{\infty}&\int^1_0{\rm d}y_1\int^{y_1}_0{\rm d}y_2\cdots\int^{y_{n-1}}_0{\rm d}y_n\,\big(\re^{-y_1\dsA}\,\dsB(s)\,\re^{y_1\dsA}\big)\non\\
	&\qquad\quad\times\big(\re^{-y_2\dsA}\,\dsB(s)\,\re^{y_2\dsA}\big)\times\cdots\times\big(\re^{-y_n\dsA}\,\dsB(s)\,\re^{y_n\dsA}\big)~.
\eaa

We extend this method to the non-minimal matrix operator derived from NLED quantisation in Section \ref{sct:QuantisationNLED}:
\be \label{eqn:NLEDMatrixOperatorDefn}
	\D=-L_{\a}\id_4\,\Box+\dsG^{ab}\pd_a \pd_b+\dsV^a\pd_a~,
\ee
where $\id_4$ is the $4\times 4$ identity matrix and $\dsG^{ab}$ and $\dsV^a$ are given by\footnote{Throughout the remainder of this paper we omit the subscript $B$ used to denote background fields 
in Section \ref{sct:QuantisationNLED}. All electromagnetic quantities, including first and second derivatives of the Lagrangian and the causality conditions, 
are now understood to be evaluated on the background.} 
\begin{subequations}
	\baa 
	(\dsG^{ab})^c_{\ph{c}d}&:=G^{cab}_{\ph{cab}d}=\big(L_{\a\a}F^{ca}+L_{\a\b}\~F^{ca}\big)F^b_{\ph{b}d}+\big(L_{\a\b}F^{ca}+L_{\b\b}\~F^{ca}\big)\~F^b_{\ph{b}d}~,\label{aln:dsGDefn}\\
	(\dsV^a)^b_{\ph{b}c}&:=V^{ba}_{\ph{ba}c}=-(\pd^aL_{\a})\d^b_{\ph{b}c}+(\pd_cL_{\a})\h^{ab}-(\pd^bL_{\a})\d^a_{\ph{a}c}+(\pd_dL_{\b})\ve^{bda}_{\ph{bda}c}+(\pd_dG^{bda}_{\ph{bda}c})~.\label{aln:dsVDefn}
	\eaa
\end{subequations}
For convenience in what follows, we introduce the following notation for the principal matrix term of the operator:
\be\label{eqn:dsHDefn}
	\dsH^{ab}:=-L_{\a}\h^{ab}\id_4+\dsG^{ab}~,\qquad\qquad \D=\dsH^{ab}\pd_a\pd_b+\dsV^a\pd_a~.
\ee 

In Subsection \ref{ssct:VolterraSeriesExpansion} we will use the Volterra-series approach to identify contributions to the heat kernel. In Subsections \ref{ssct:QuarticA0ContributionFlat}, \ref{ssct:QuarticA1ContributionFlat} and \ref{ssct:QuarticA2ContributionFlat} we compute the leading order terms of the $a_0$, $a_1$ and $a_2$ contributions, by expanding up to quartic order in the background field strength $\cO(F^4)$.

\subsection{Volterra series expansion} \label{ssct:VolterraSeriesExpansion}

The heat kernel $K(x,x';s)$ associated with the NLED operator $\D$ (\ref{eqn:NLEDMatrixOperatorDefn}) is defined by
\be
	K(x,x';s)=\re^{\ri s \D}\,\d^{(4)}(x-x')\,\id_4~.
\ee
Using a Fourier integral representation of the delta function,\footnote{Strictly speaking, the delta function should be accompanied by an additional factor $\cI(x,x')$ satisfying $\cI(x,x) = \id_4$ (for instance, the parallel displacement propagator) to ensure the correct gauge transformation properties of the heat kernel. However, this factor may be safely omitted when computing diagonal heat kernel coefficients in a flat background spacetime.}
\be \label{eqn:IntegralDeltaFlat}
	\d^{(4)}(x-x')=\inte k \, \re^{\ri k_a(x^a-x'^a)}~,\qquad \rd k:=\frac{\rd^4 k}{(2\p)^4}~,
\ee
we can act through with the operator, and the heat kernel becomes
\be\label{eqn:Non-CoincidenceLimitHeatKernelFlat}
	K(x,x';s)=\inte k \, \re^{\ri s \D} \re^{\ri k_a(x^a-x'^a)} = \inte k \, \re^{\ri k_a(x^a-x'^a)}\re^{\ri s \hat{\D}}~,
\ee
with the shifted operator
\be 
	\hat{\D}:=\dsH^{ab}X_aX_b+\dsV^aX_a~,\qquad\qquad X_a:=\pd_a+\ri k_a~.
\ee
To compute the one-loop effective action (\ref{eqn:HeatKernelEARelash}), it is sufficient to consider only the functional trace of the heat kernel
\be
	K(s)=\inte^4x\,K(x;s)~,\qquad K(x;s):=\lim_{x'\rightarrow x} \tr\,K(x,x';s)\,,
\ee
in which case (\ref{eqn:Non-CoincidenceLimitHeatKernelFlat}) becomes
\be \label{eqn:CoincidenceLimitHeatKernelFlat}
	K(x;s)=\tr\inte k\,\re^{\ri s \hat{\D}}~.
\ee
For the purposes of a derivative expansion and identification of DeWitt coefficients, it is useful to make the $k$-dependence explicit in $\hat{\D}$
\be \label{eqn:dsPDefn}
	\hat{\D}=-k_ak_b\dsH^{ab}+k_a\dsP^a+\D~,\qquad 
	\dsP^a:=2\ri\dsH^{(ab)} \pd_b +\ri\dsV^a~.
\ee
Rescaling the momentum integration (\ref{eqn:CoincidenceLimitHeatKernelFlat}) by $k_a\rightarrow k_a/\sqrt{s}$, allows us to recast the exponent in powers of proper time $s$
\be
	K(x;s)=\frac{\tr}{s^2}\inte k\,\re^{-\ri k_ak_b\dsH^{ab}+\ri \sqrt{s}\,k_a\dsP^a+\ri s \D}~.
\ee
We can extract the $s$-dependence from the exponent by choosing 
\be \label{eqn:AandBDefnFlat}
	\dsA:=-\ri k_ak_b\dsH^{ab}~,\qquad \dsB(s):=\ri\sqrt{s}\,k_a\dsP^a+\ri s\D~,
\ee
within the following Volterra identity
\baa 
	\re^{\dsA+\dsB(s)}=\re^\dsA+\re^\dsA\sum_{n=1}^{\infty}&\int^1_0{\rm d}y_1\int^{y_1}_0{\rm d}y_2\cdots\int^{y_{n-1}}_0{\rm d}y_n\,\big(\re^{-y_1\dsA}\,\dsB(s)\,\re^{y_1\dsA}\big)\non\\
	&\qquad\quad\times\big(\re^{-y_2\dsA}\,\dsB(s)\,\re^{y_2\dsA}\big)\times\cdots\times\big(\re^{-y_n\dsA}\,\dsB(s)\,\re^{y_n\dsA}\big)~.
\eaa
Following the notation of \cite{Iochum:2016ynh,Iochum:2017ver,Grasso:2023qye}, for arbitrary matrices $\dsQ_1,\ldots,\dsQ_n$, we define
\baa \label{eqn:DefnofHIntegralsFlat}
	H_n[\dsQ_1\otimes \dsQ_2\otimes\cdots\otimes \dsQ_n]:=\re^\dsA&\int^1_0{\rm d}y_1\int^{y_1}_0{\rm d}y_2\cdots\int^{y_{n-1}}_0{\rm d}y_n\,\big(\re^{-y_1\dsA}\,\dsQ_1\,\re^{y_1\dsA}\big)\non\\
	&\qquad\times\big(\re^{-y_2\dsA}\,\dsQ_2\,\re^{y_2\dsA}\big)\times\cdots\times\big(\re^{-y_n\dsA}\,\dsQ_n\,\re^{y_n\dsA}\big)~.
\eaa
The heat kernel is realised in this notation as follows:
\be 
	K(x;s)=\frac{\tr}{s^2}\inte k\,\bigg\{\re^\dsA+\sum_{n=1}^{\infty}H_n[\dsB(s)\otimes \dsB(s)\otimes\cdots\otimes \dsB(s)]\bigg\}~.
\ee
To identify the DeWitt coefficient contributions to the heat kernel, we expand $\dsB(s)$ and consider terms up to and including order $s^0$:
\baa \label{aln:VolterraSeriesResult}
	K(x;s)&=\frac{\tr}{s^2}\inte k\,\re^{-\ri k_ak_b\dsH^{ab}}+\frac{\tr}{s}\inte k\,\bigg\{\ri H_1[\D]-k_ak_bH_2[\dsP^a\otimes\dsP^b]\bigg\}\non\\
	&+\tr\inte k\,\biggl\{-H_2[\D\otimes\D]-\ri k_ak_b\bigg(H_3[\D\otimes\dsP^a\otimes\dsP^b]+H_3[\dsP^a\otimes\D\otimes\dsP^b]\non\\
	&\qquad\qquad\quad+H_3[\dsP^a\otimes\dsP^b\otimes\D]\bigg)+k_ak_bk_ck_dH_4[\dsP^a\otimes\dsP^b\otimes\dsP^c\otimes\dsP^d]\biggr\}\non\\
	&+\cO(s)~.
\eaa
Comparison of the above expression with the asymptotic expansion of $K(x;s)$ (\ref{eqn:DeWittCoefficientsFlat}), we can directly identify the DeWitt coefficients. In particular, the above result includes the $a_0$, $a_1$ and $a_2$ contributions. As anticipated, no fractional powers of $s$ appear in the above series, since such terms are odd in $k$ and therefore vanish under the integral.

It is important to note that the Volterra-series expansion naturally yields a derivative expansion for the NLED operator. In this context, we count the derivatives already acting on, or still available to act on, the background field strengths. The only sources of derivative contributions come from the operators $\D$ and $\dsP^a$. At first sight, the operator $\D$ (\ref{eqn:NLEDMatrixOperatorDefn}) does not appear to have definite order in derivatives. However, its principal term $\dsH^{ab}\pd_a\pd_b$ and linear term $\dsV^a\pd_a$ are both second order in derivatives. This follows from the definition of $\dsV^a$ (\ref{aln:dsVDefn}), which already contains a single derivative acting on the background field strength. Thus, $\D$ contributes second order in derivatives. From similar arguments, $\dsP^a$ (\ref{eqn:dsPDefn}) proves to be linear in derivatives. Consequently, the $a_n$ contribution to the heat kernel is fixed to be order $2^n$ in derivatives of the background field strength.


\subsection{The quartic order $a_0$ contribution}\label{ssct:QuarticA0ContributionFlat}

In this subsection, we study the $a_0$ contribution to the heat kernel, denoted $K(x;s)\big|_{a_0}$. This piece contains no derivatives on the background field strength. It is given by
\baa \label{eqn:IdentifyA0Contri}
	K(x;s)\Big|_{a_0}=\frac{\tr}{s^2}\inte k\,\re^{\dsA}~,\qquad\qquad \dsA=-\ri k_ak_b \dsH^{ab}~.
\eaa
In this work, we consider a generic NLED theory analytic around zero in the weak-field regime.\footnote{With the exception of Maxwell electrodynamics, conformal NLED theories are non-analytic around zero and therefore excluded from this approach. We will consider this class of theories in Section \ref{sct:ConformalHeatKernelTechniques}.} Since the Lagrangian is a scalar function of the two background Lorentz invariants $\a$ and $\b$, it admits an expansion in even powers of the background field strength
\baa
L(\a,\b)&=\cancelto{0}{L(0,0)}+L_{\a}\big|_{\a=\b=0}~\a+L_\b\big|_{\a=\b=0}~\b\non\\
&\qquad+\frac{1}{2}L_{\a\a}\big|_{\a=\b=0}~\a^2+L_{\a\b}\big|_{\a=\b=0}~\a\b+\frac{1}{2}L_{\b\b}\big|_{\a=\b=0}~\b^2+\cO(F^6)~.
\eaa
The expansion in even powers is also inherited by any derivatives of the Lagrangian with respect to $\a$ and $\b$. This property proves to cascade into all composite fields introduced in earlier sections, ensuring that they too admit expansions in even powers. In particular, the $a_0$, $a_1$ and $a_2$ contributions to the heat kernel share this property. For brevity, we will use a superscript $(n)$ to denote the order in background field strength, e.g.
\be
\left\{
\begin{array}{l}
	Q^{(0)} ~ \text{--- constant contribution,} \\[0.5em]
	Q^{(2)} ~ \text{--- quadratic contribution,} \\[0.5em]
	Q^{(4)} ~ \text{--- quartic contribution.}
\end{array}
\right.\non
\ee
From the definitions of $\dsG^{ab}$ (\ref{aln:dsGDefn}) and $\dsV^a$ (\ref{aln:dsVDefn}), it follows that neither have constant contributions:
\begin{subequations}
	\baa
	\dsG^{ab}&=\dsG^{ab(2)}+\dsG^{ab(4)}+\cO(F^6)~,\label{aln:dsGQuadraticOrder}\\
	\dsV^{a}&=\dsV^{a(2)}+\dsV^{a(4)}+\cO(F^6)~.
	\eaa
\end{subequations}

Let us now expand the $a_0$ contribution \eqref{eqn:IdentifyA0Contri} up to quartic order in the background field strength
\baa 
	K(x;s)\Big|_{a_0}&=\frac{\tr}{s^2}\inte k\,\re^{\dsA^{(0)}+\dsA^{(2)}+\dsA^{(4)}}+\cO(F^5)\non\\
	&=\frac{\tr}{s^2}\inte k\,\re^{\ri k^2 L_{\a}^{(0)}}\big[\id_4-\ri k_ak_b\dsH^{ab(2)}-\ri k_ak_b\dsH^{ab(4)}-k_ak_bk_ck_d\dsH^{ab(2)}\dsH^{cd(2)}\big]+\cO(F^5)~,
\eaa
where we have introduced the notation $k^2=k^ak_a$. The momentum space integrals can be computed using Gaussian moments 
\be \label{eqn:a0GaussianMomentIdentity}
	\inte k\,k_{a_1}k_{a_2}\cdots k_{a_{2n}}\re^{\ri k^2 		L_{\a}^{(0)}}=\bigg(\frac{\ri}{2L_{\a}^{(0)}}\bigg)^n\frac{h}{(L_{\a}^{(0)})^2}~\h_{a_1a_2\ldots a_{2n}}~, \quad h=\frac{\ri}{(4\p \ri)^2}~,
\ee
where we choose the sign convention $L_{\a}^{(0)}<0$ and denote $\h_{a_1a_2\ldots a_{2n}}$ as the independent terms in the symmetrisation\footnote{In this work, symmetrisation of $n$ indices includes a $1/n!$ factor.} of $n$ Minkowski metric products
\be
	\h_{a_1a_2\ldots a_{2n}}:=2^n n!~\h_{(a_1a_2}\h_{a_3a_4}\cdots \h_{a_{2n-1}a_{2n})}~,
\ee
which can also be obtained recursively using
\be 
	\h_{a_1a_2\ldots a_{2n}}=\h_{a_1a_2}\h_{a_3a_4\ldots a_{2n}}+\h_{a_1a_3}\h_{a_2a_4\ldots a_{2n}}+\cdots+\h_{a_1a_{2n}}\h_{a_2a_3\ldots a_{2n-1}}~.
\ee
The expression \eqref{eqn:a0GaussianMomentIdentity} can be proven inductively using a Gaussian moment generating identity
\be 
	0=\inte k\,\frac{\pd}{\pd k_b}\Big(k_{a_1}k_{a_2}\cdots k_{a_{2n-1}}\re^{\ri k^2 L_{\a}^{(0)}}\Big)~.
\ee
After evaluating the momentum space integrals and comparing with the asymptotic expansion (\ref{eqn:DeWittCoefficientsFlat}), we find the quartic order $a_0$ DeWitt coefficient is given by
\baa\label{aln:NLEDa0HResult}
	a_0&=\frac{4}{(L_{\a}^{(0)})^2}+\frac{2}{(L_{\a}^{(0)})^3}~\tr\Big[\dsH^{a\ph{a}(2)}_{\ph{a}a}+\dsH^{a\ph{a}(4)}_{\ph{a}a}\Big]\non\\
	&\qquad+\frac{1}{4(L_{\a}^{(0)})^4}~\tr\Big[\dsH^{a\ph{a}(2)}_{\ph{a}a}\dsH^{b\ph{b}(2)}_{\ph{b}b}+2\dsH^{ab(2)}\dsH^{(2)}_{(ab)}\Big]+\cO(F^5)~.
\eaa
Since $\dsH^{ab(2)}$ and $\dsH^{ab(4)}$ are composite fields \eqref{eqn:dsHDefn}, we can expand further
\baa\label{aln:NLEDa0GResult}
	a_0&=\frac{4}{(L_{\a}^{(0)})^2}+\frac{2}{(L_{\a}^{(0)})^3}\Bigl[-16L_{\a}^{(2)}+G^{ab~(2)}_{\ph{ab}ba}-16L_{\a}^{(4)}+G^{ab~(4)}_{\ph{ab}ba}\Bigr]\non\\
	&\quad+\frac{1}{4(L_{\a}^{(0)})^4}\Big[96\big(L_{\a}^{(2)}\big)^2-12L_{\a}^{(2)}G^{ab~(2)}_{\ph{ab}ba}+G^{ab~(2)}_{\ph{ab}bc}G^{cd~(2)}_{\ph{cd}da}+2G^{abcd(2)}G^{~(2)}_{d(bc)a}\Big]+\cO(F^5)~.
\eaa
Eqs.~\eqref{aln:NLEDa0HResult} and~\eqref{aln:NLEDa0GResult} represent the general result for the $a_0$ coefficient to quartic order for an arbitrary NLED model. 

As an example, let us now state the result for the Born--Infeld theory. In this case, we have constant contributions given by:
\be 
	L_{\a}^{(0)}=-1~,\qquad L_{\a\a}^{(0)}=L_{\b\b}^{(0)}=\frac{1}{T}~,\qquad L_{\a\b}^{(0)}=0~,
\ee
as well as quadratic and quartic contributions:
\be 
	L_{\a}^{(2)}=\frac{\a}{T}~, \quad L_{\a\a}^{(2)}=-\frac{3\a}{T^2}~, \quad L_{\a\b}^{(2)}=-\frac{\b}{T^2}~,\quad L_{\b\b}^{(2)}=-\frac{\a}{T^2}~, \quad L_{\a}^{(4)}=-\frac{3\a^2}{2T^2}-\frac{\b^2}{2T^2}~.
\ee
Substituting this into (\ref{aln:NLEDa0GResult}) yields
\be
	(a_0)_{\rm BI}=4+\frac{8\a}{T}+\frac{4}{T^2}\big(6\a^2-\b^2\big)+\cO(F^5)~.
\ee


\subsection{The quartic order $a_1$ contribution}\label{ssct:QuarticA1ContributionFlat}

In this subsection we study the $a_1$ contribution to the heat kernel, denoted $K(x;s)\big|_{a_1}$. This piece contains two derivatives free to act on or already acting on the background field strength. It is given by:
\baa \label{eqn:IdentifyA1ContriFlat}
	K(x;s)\Big|_{a_1}&=\frac{\tr}{s}\inte k\,\bigg\{\ri H_1[\D]-k_ak_bH_2[\dsP^a\otimes\dsP^b]\bigg\}~.
\eaa
One can show that for the NLED operator, the momentum space integrals of $H_n$ cannot be evaluated directly with differential operators in 
their arguments.\footnote{As was done in  \cite{Grasso:2023qye} for the $a_1$ contribution to the heat kernel.} To resolve this, we will use a series of manipulations to recast 
the above expression into computable Gaussian integrals. To make progress, we will concentrate on the leading order contribution to $a_1$ which proves to be quartic order in the background field strength.

First, let us outline the steps required to compute the quartic order $a_1$ contribution:
\begin{enumerate}[label=(\roman*), itemsep=1mm, topsep=1mm, parsep=1mm]
	\item For a given choice of derivative splitting (see below for a detailed explanation), move all free derivatives originally appearing in the arguments of $H_n$ to either the left or the right.
	\item Identify all contributions to $a_1$ up to and including quartic order $\cO(F^4)$.
	\item Evaluate the $y$-integrals present in the $H_n$ expressions.
	\item Evaluate the outstanding momentum space integrals.
	\item Express all resulting structures in terms of a chosen set of basis elements which are unique up to matrix trace and integration by parts.
\end{enumerate}

To achieve (i), we note that the following relation holds for an arbitrary parameter $\l$ and matrix operator $\dsO(x)$
\be 
	\big(\pd_a\re^{\l \dsO}\big)=\re^{\l \dsO}\int^\l_0\rd \t\,\re^{-\t \dsO}\big(\pd_a \dsO\big)\re^{\t \dsO}~,
\ee
known as Duhamel's formula. The identity above serves as the foundation for a proof within \cite{Iochum:2016ynh} that demonstrates how to move free derivatives within the arguments of $H_n$ to the right. Specifically, this operation yields\footnote{Note that the coincidence limit heat kernel $K(x;s)=K(x;s)\cdot1$ is not a differential operator, so any $H_n$ terms with free derivatives at the end are identically zero, for example $H_n[\dsQ_1\otimes \dsQ_2\otimes\cdots\otimes \dsQ_n]\pd_a$.}
\baa \label{aln:PushingDerivativesRight}
	H_n[\dsQ_1&\otimes \dsQ_2\otimes \cdots \otimes \dsQ_{i}\pd_a\otimes \dsQ_{i+1}\otimes \cdots \otimes 
	\dsQ_n]=\sum_{j=i+1}^nH_n[\dsQ_1\otimes \dsQ_2 \otimes (\pd_a\dsQ_j)\otimes\cdots\otimes \dsQ_n]\non\\
	& +\sum_{j=i}^{n}H_{n+1}[\dsQ_1\otimes \dsQ_2 \otimes \cdots \otimes \dsQ_{j} \otimes (\pd_a\dsA) \otimes \dsQ_{j+1} \otimes \cdots \otimes \dsQ_n]~,
\eaa
where $\dsQ_1,\dsQ_2,\ldots \dsQ_n$ are arbitrary matrices and the matrix $\dsA:=-\ri k_ak_b\dsH^{ab}$ defined in (\ref{eqn:AandBDefnFlat}). A corollary of the above \cite{Grasso:2023qye} allows us to pull free derivatives to the left, which accomodates integration by parts\footnote{In this work we ignore all boundary terms.}
\baa \label{aln:PushingDerivativesLeft}
	&H_n[\dsQ_1\otimes \dsQ_2\otimes \cdots \otimes \dsQ_{i}\pd_a\otimes \dsQ_{i+1}\otimes \cdots \otimes \dsQ_n]\non\\
	&=\pd_a(H_n[\dsQ_1\otimes \dsQ_2\otimes\cdots\otimes \dsQ_n])-\sum_{j=1}^iH_n[\dsQ_1\otimes \dsQ_2 \otimes (\pd_a\dsQ_j)\otimes\cdots\otimes \dsQ_n]\non\\
	&\qquad-\sum_{j=1}^{i}H_{n}[\dsQ_1\otimes \dsQ_2 \otimes \cdots \otimes \dsQ_{j} \otimes (\pd_a\dsA) \otimes \dsQ_{j+1} \otimes \cdots \otimes \dsQ_n]~.
\eaa

For each free derivative, we have the freedom to split $\pd_a=\r\,\pd_a+(1-\r)\,\pd_a$, with $\r$ some arbitrary parameter to be fixed later. The first term is pulled entirely to the left using integration by parts; the second pushed completely to the right.  In this work, we consider three choices of splitting the derivatives:
\begin{enumerate}[label=(\alph*), itemsep=1mm, topsep=1mm, parsep=1mm]
	\item \underline{$\r = 0$}: Push all free derivatives $\pd_a$ completely to the right.
	\item \underline{$\r = \frac{1}{2}$}: Split free derivatives in two. Push half to the right and half to the left.
	\item \underline{Minimal derivative prescription}: Spread the free derivatives asymmetrically between the left and  right, so the arguments of $H_n$ contain no more than one derivative acting on the background field strength.
\end{enumerate}
Our results agree for all three prescriptions, modulo total derivative terms. After moving all derivatives to either side, the terms remaining contain no differential operators in their arguments, and are purely background matrices. 

To compute the approximation to $a_1$ for step (ii), we consider a generic NLED theory analytic around zero in the weak-field regime and expand the background field strength up to and including quartic order $\cO(F^4)$. As was described in Subsection \ref{ssct:QuarticA0ContributionFlat} this results in an $a_1$ contribution of the heat kernel which admits an expansion in even powers of the background field strength. The $a_1$ DeWitt coefficient proves to have no constant contributions, and the quadratic contributions which are total derivatives, make no contribution. Therefore, the leading term in the weak-field expansion is of quartic order. Recall, for brevity, we introduced a superscript $(n)$ to denote order in background field strength:
\be
\left\{
\begin{array}{l}
	Q^{(0)} ~ \text{--- constant contribution,} \\[0.5em]
	Q^{(2)} ~ \text{--- quadratic contribution,} \\[0.5em]
	Q^{(4)} ~ \text{--- quartic contribution.}
\end{array}
\right.\non
\ee

Following steps (i) and (ii), we are now left with a collection of $H_n$ with arguments
\be
	H_n[\dsQ_1\otimes \dsQ_2\otimes\cdots\otimes \dsQ_n]~,
\ee
where the $\dsQ_i$ are known matrices, which may carry additional Lorentz indices. Recalling the general definition (\ref{eqn:DefnofHIntegralsFlat})
\baa\label{eqn:GeneralDefnofHInts}
	H_n[\dsQ_1\otimes \dsQ_2\otimes\cdots\otimes \dsQ_n]:=\re^\dsA&\int^1_0{\rm d}y_1\int^{y_1}_0{\rm d}y_2\cdots\int^{y_{n-1}}_0{\rm d}y_n\,\big(\re^{-y_1\dsA}\,\dsQ_1\,\re^{y_1\dsA}\big)\non\\
	&\times\big(\re^{-y_2\dsA}\,\dsQ_2\,\re^{y_2\dsA}\big)\times\cdots\times\big(\re^{-y_n\dsA}\,\dsQ_n\,\re^{y_n\dsA}\big)~,
\eaa
our goal in step (iii) is to evaluate the $y$-integrals present within the above definition of $H_n$. In the above expression each factor within a round bracket may be simplified since the $\dsQ_i$ are no longer differential operators and we can expand $\dsA=\dsA^{(0)}+\dsA^{(2)}+ \cdots$
\baa 
	\big(\re^{-y_i\dsA}\,\dsQ_i\,\re^{y_i\dsA}\big)&=\big(\re^{-y_i (\ri k^2L_\a^{(0)}\id_4+\dsA^{(2)})}\,\dsQ_i\,\re^{y_i (\ri k^2L_\a^{(0)}\id_4+\dsA^{(2)})}\big)+\cdots\non\\
	&=\big(\re^{-y_i \dsA^{(2)}}\,\dsQ_i\,\re^{y_i \dsA^{(2)}}\big)+\cdots\non\\
	&= \dsQ_i+\big[\dsQ_i,\dsA^{(2)}\big]y_i+\cdots~,
\eaa
where we have noted that the tensor field $\dsG^{ab}$ is at least quadratic order or higher (\ref{aln:dsGQuadraticOrder}), so does not contribute to $\dsA^{(0)}$. It follows that an arbitrary $H_n$ can be expanded into contributions as
\be
	H_n[\dsQ_1\otimes \dsQ_2\otimes\cdots\otimes \dsQ_n]=H_n^{(0)}[\dsQ_1\otimes \dsQ_2\otimes\cdots\otimes \dsQ_n]+H_n^{(2)}[\dsQ_1\otimes \dsQ_2\otimes\cdots\otimes \dsQ_n]+~\cdots~,
\ee
where here the superscript denotes the overall field strength contribution to $H_n$ coming solely from $\dsA=\dsA^{(0)}+\dsA^{(2)}+ \cdots$ appearing in the definition of $H_n$. For example, one can show:
\be \label{eqn:H_nConstant}
	H_n^{(0)}[\dsQ_1\otimes \dsQ_2\otimes\cdots\otimes \dsQ_n]=\frac{\re^{\ri k^2 L_{\a}^{(0)}}}{n!}~\dsQ_1\dsQ_2\cdots\dsQ_n~,
\ee
and
\baa \label{eqn:H_nQuadratic}
	H_n^{(2)}[\dsQ_1\otimes \dsQ_2\otimes\cdots\otimes \dsQ_n]=\frac{\re^{\ri k^2 L_{\a}^{(0)}}}{(n+1)!}\{\dsA^{(2)}\dsQ_1&\dsQ_2\cdots\dsQ_n+\dsQ_1\dsA^{(2)}\dsQ_2\cdots\dsQ_n\non\\
	&+\cdots+\dsQ_1\dsQ_2\cdots\dsQ_n\dsA^{(2)}\}~.
\eaa 

For step (iv), the outstanding momentum space integrals can be computed using the Gaussian moments given in equation \eqref{eqn:a0GaussianMomentIdentity}. All that remains is step (v), collecting all results together and simplifying by expressing everything in terms of a set of Lorentz scalar basis structures, unique up to matrix trace and integration by parts.

Our remarks so far have been quite general, so let us illustrate the process on the first integral in (\ref{eqn:IdentifyA1ContriFlat})
\be \label{eqn:a1ExampleCalc}
	K(x;s)\Big|_{a_1,\,\D}:=\frac{\ri\,\tr}{s}\inte k\, H_1[\D]=\frac{\ri\,\tr}{s}\inte k\, \bigg\{H_1[\dsV^a\pd_a]+H_1[\dsH^{ab}\pd_a\pd_b]\bigg\}~.
\ee
First, consider pushing all the derivatives to the right according to (\ref{aln:PushingDerivativesRight}), leaving no free derivatives within the arguments of $H_n$.  We obtain
\baa
	H_1[\D]&= H_2[\dsV^a\otimes(\pd_a\dsA)]+H_2[\dsH^{ab}\otimes(\pd_a\pd_b\dsA)]\non\\
	&\qquad+H_3[\dsH^{ab}\otimes(\pd_a\dsA)\otimes(\pd_b\dsA)]+H_3[\dsH^{ab}\otimes(\pd_b\dsA)\otimes(\pd_a\dsA)]~.
\eaa
Extracting the $k$-dependence from $\dsA$ and retaining terms up to quartic order we find the following result
\baa 
	&H_1[\D]=-\ri k_{a_1}k_{a_2}\bigg(H_2^{(0)}[\dsV^{a(2)}\otimes(\pd_a\dsH^{a_1a_2(2)})]+H_2^{(0)}[\dsH^{ab(0)}\otimes(\pd_a\pd_b\dsH^{a_1a_2(2)})]\non\\
	&\qquad\qquad\qquad\qquad +H_2^{(0)}[\dsH^{ab(2)}\otimes(\pd_a\pd_b\dsH^{a_1a_2(2)})]+H_2^{(0)}[\dsH^{ab(0)}\otimes(\pd_a\pd_b\dsH^{a_1a_2(4)})]\non\\
	&\qquad\qquad\qquad\qquad +H_2^{(2)}[\dsH^{ab(0)}\otimes(\pd_a\pd_b\dsH^{a_1a_2(2)})]\bigg)\non\\
	&- k_{a_1}k_{a_2}k_{a_3}k_{a_4}\bigg(
	H_3^{(0)}[\dsH^{ab(0)}\otimes(\pd_a\dsH^{a_1a_2(2)})\otimes(\pd_b\dsH^{a_3a_4(2)})]\non\\
	&\qquad\qquad\qquad\quad+H_3^{(0)}[\dsH^{ab(0)}\otimes(\pd_b\dsH^{a_1a_2(2)})\otimes(\pd_a\dsH^{a_3a_4(2)})]\bigg)+\cO(F^5)~.
\eaa
Evaluating the $y$-integrals present in $H_n^{(0)}$ and $H_n^{(2)}$ using (\ref{eqn:H_nConstant}) and (\ref{eqn:H_nQuadratic}) and inserting into the overall 
momentum space integral (\ref{eqn:a1ExampleCalc}) we obtain
\baa
	K(x;s)&\Big|_{a_1,\,\D}=\frac{1}{2s}\,\tr\Big[\dsV^{a(2)}(\pd_a\dsH^{a_1a_2(2)})+\dsH^{ab(0)}(\pd_a\pd_b\dsH^{a_1a_2(2)})\non\\
	&\qquad\qquad\qquad+\dsH^{ab(2)}(\pd_a\pd_b\dsH^{a_1a_2(2)})+\dsH^{ab(0)}(\pd_a\pd_b\dsH^{a_1a_2(4)})\Big]\non\\
	&\qquad\qquad\qquad\qquad\times\inte k\,k_{a_1}k_{a_2}\re^{\ri k^2 L_{\a}^{(0)}}\non\\
	&-\frac{\ri}{3!s}\,\tr\Big[2\dsH^{a_3a_4(2)}\dsH^{ab(0)}(\pd_a\pd_b\dsH^{a_1a_2(2)})+\dsH^{ab(0)}\dsH^{a_3a_4(2)}(\pd_a\pd_b\dsH^{a_1a_2(2)})\non\\
	&\qquad\qquad+\dsH^{ab(0)}(\pd_a\dsH^{a_1a_2(2)})(\pd_b\dsH^{a_3a_4(2)})+\dsH^{ab(0)}(\pd_b\dsH^{a_1a_2(2)})(\pd_a\dsH^{a_3a_4(2)})\Big]\non\\
	&\qquad\qquad\qquad\times\inte k\,k_{a_1}k_{a_2}k_{a_3}k_{a_4}\re^{\ri k^2 L_{\a}^{(0)}}+\cO(F^5)~.
\eaa
The momentum space integrals are evaluated using   (\ref{eqn:a0GaussianMomentIdentity}). Finally, we identify a set of basis structures and simplify the result modulo total derivatives terms. We find the $a_1$ contribution from this integral is
\baa
	K(x;s)\Big|_{a_1,\,\D}=\frac{\ri h}{s\big(L_{\a}^{(0)}\big)^3}~\tr\bigg[&\frac{1}{4}\dsV_a^{(2)}\big(\pd^a\dsH^{b\ph{b}(2)}_{\ph{b}b}\big)+\frac{1}{24}\big(\pd^a\dsH^{b\ph{b}(2)}_{\ph{b}b}\big)\big(\pd_a\dsH^{c\ph{c}(2)}_{\ph{c}c}\big)\non\\
	&-\frac{1}{4}\big(\pd^a\dsH_{ab}^{(2)}\big)\big(\pd^b\dsH^{c\ph{c}(2)}_{\ph{c}c}\big)+\frac{1}{12}\big(\pd^a\dsH^{bc(2)}\big)\big(\pd_a\dsH^{(2)}_{(bc)}\big)\bigg]+\cO(F^5)~.
\eaa
We repeat the above computation for the second integral in (\ref{eqn:IdentifyA1ContriFlat}). Combining the results, we find that the leading contribution to $a_1$ for a generic NLED theory 
expanded in the weak-field regime is
\baa
	a_1&=\frac{1}{\big(L_{\a}^{(0)}\big)^3}~\tr\bigg[\frac{1}{4}\dsV^{a(2)}\dsV_a^{(2)}+\frac{1}{4}\dsV_a^{(2)}\big(\pd^a\dsH^{b\ph{b}(2)}_{\ph{b}b}\big)-\frac{1}{2}\dsV_a^{(2)}\big(\pd_b\dsH^{(ab)(2)}\big)+\frac{1}{48}\big(\pd^a\dsH^{b\ph{b}(2)}_{\ph{b}b}\big)\big(\pd_a\dsH^{c\ph{c}(2)}_{\ph{c}c}\big)\non\\
	&\qquad\qquad\qquad-\frac{1}{6}\big(\pd^a\dsH_{ab}^{(2)}\big)\big(\pd^b\dsH^{c\ph{c}(2)}_{\ph{c}c}\big)+\frac{1}{24}\big(\pd^a\dsH^{bc(2)}\big)\big(\pd_a\dsH^{(2)}_{(bc)}\big)+\frac{1}{12}\big(\pd^a\dsH_{(ab)}^{(2)}\big)\big(\pd^c\dsH^{b\ph{c}(2)}_{\ph{b}c}\big)\non\\
	&\qquad\qquad\qquad+\frac{1}{12}\big(\pd^a\dsH_{(ab)}^{(2)}\big)\big(\pd^c\dsH^{\ph{c}b(2)}_c\big)\bigg]+\cO(F^5)~.
	\label{e10}
\eaa
This result is consistent with other approaches.  For example, if we set $L_\a=-1$ and $\dsG_{ab}=0$ which corresponds to Maxwell theory, the operator \eqref{eqn:NLEDMatrixOperatorDefn} becomes minimal, and we recover the quartic terms generated by the Schwinger--DeWitt approach. Similarly, agreement with the results of~\cite{Grasso:2023qye} 
is obtained by solely setting $\dsG_{ab}=0$ and expanding up to and including quartic order.

Since $\dsH_{ab}^{(2)}$ and $\dsV_a^{(2)}$ are composite fields, we can further decompose our result in terms of $L_{\a}^{(2)}$, $L_{\b}^{(2)}$ and $G_{abcd}^{(2)}$, which yields
\baa\label{aln:NLEDa1GResult}
	a_1=\frac{1}{\big(L_{\a}^{(0)}\big)^3}\bigg[&\frac{3}{2}\big(\pd^a L_{\a}^{(2)}\big)\big(\pd_a L_{\a}^{(2)}\big)+\frac{3}{2}\big(\pd^a L_{\b}^{(2)}\big)\big(\pd_a L_{\b}^{(2)}\big)-\frac{1}{3}\big(\pd^a L_{\a}^{(2)}\big)\big(\pd_a G^{bc~(2)}_{\ph{bc}cb}\big)\non\\
	&+\frac{1}{3}\big(\pd^a L_{\a}^{(2)}\big)\big(\pd_b G^{bc~(2)}_{\ph{bc}ca}\big)-\frac{1}{2}\,\ve^{acde}\big(\pd_a L_{\b}^{(2)}\big)\big(\pd^b G^{(2)}_{bcde}\big)+\frac{1}{48}\big(\pd^a G_{b\ph{c}cd}^{\ph{b}c(2)}\big)\big(\pd_a G^{de\ph{e}b(2)}_{\ph{de}e}\big)\non\\
	&+\frac{1}{24}\big(\pd^a G^{bcde(2)}\big)\big(\pd_a G_{e(cd)b}^{(2)}\big)+\frac{1}{12}\big(\pd^aG^{(2)}_{bacd}\big)\big(\pd^cG^{de\ph{e}b(2)}_{\ph{de}e}\big)\non\\
	&-\frac{1}{6}\big(\pd^aG_{bacd}^{(2)}\big)\big(\pd_eG^{decb(2)}\big)+\frac{1}{12}\big(\pd^aG_{bacd}^{(2)}\big)\big(\pd_eG^{dceb(2)}\big)\bigg]+\cO(F^5)~.
\eaa
Eqs.~\eqref{e10} and~\eqref{aln:NLEDa1GResult} represent the general result for the $a_1$ coefficient to leading order for an arbitrary NLED model. 

As an example, let us now state the  result for Born--Infeld theory.
In this case we have
\be 
	L_{\a}^{(0)}=-1~,\qquad L_{\a\a}^{(0)}=L_{\b\b}^{(0)}=\frac{1}{T}~,\qquad L_{\b}^{(0)}=L_{\a\b}^{(0)}=0~,
\ee
and the quadratic contributions 
\be 
	L_{\a}^{(2)}=\frac{\a}{T}~,\qquad L_{\b}^{(2)}=\frac{\b}{T}~.
\ee
Substituting this into (\ref{aln:NLEDa1GResult}) yields
\baa
	(a_1)_{\rm BI}=\frac{1}{T^2}\biggl[&\frac{5}{6}\big(\pd^a \a\big)\big(\pd_a \a \big)-2\big(\pd^a \a\big)\big(\pd_a \b\big)+\frac{1}{2}\big(\pd^a \b\big)\big(\pd_a \b\big)-\big(\pd^a\b\big)\,\pd^b(F^2)_{ab}\non\\
	&+\frac{1}{4}\,\pd^a(F^2)^{bc}\,\pd_a(F^2)_{bc}-\frac{1}{3}\,\pd^a(F^2)_{ab}\,\pd^c(F^2)^b_{\ph{b}c}\bigg]+\cO(F^5)~.
\eaa


\subsection{The quartic order $a_2$ contribution}\label{ssct:QuarticA2ContributionFlat}

In this subsection, we study the $a_2$ contribution to the heat kernel, denoted $K(x;s)\big|_{a_2}$. This piece contains four derivatives free to act on or already acting on the background field strength. It is given by:
\baa \label{eqn:IdentifyA2ContriFlat}
	K(x;s)\Big|_{a_2}&=\tr\inte k\,\bigg\{-H_2[\D\otimes\D]-\ri k_ak_b\bigg(H_3[\D\otimes\dsP^a\otimes\dsP^b]+H_3[\dsP^a\otimes\D\otimes\dsP^b]\non\\
	&\qquad\qquad\quad+H_3[\dsP^a\otimes\dsP^b\otimes\D]\bigg)+k_ak_bk_ck_dH_4[\dsP^a\otimes\dsP^b\otimes\dsP^c\otimes\dsP^d]\bigg\}~.
\eaa

As described in Subsection \ref{ssct:QuarticA0ContributionFlat}, when computing quartic order DeWitt coefficients we consider a generic NLED analytic around zero. This results in an $a_2$ contribution of the heat kernel which admits an expansion in even powers of the background field strength. 

With the groundwork already laid in Subsection \ref{ssct:QuarticA1ContributionFlat}, the algorithm to compute the quartic contribution to the induced action is as follows:
\begin{enumerate}[label=(\roman*), itemsep=1mm, topsep=1mm, parsep=1mm]
	\item For a given choice of derivative splitting, move all free derivatives originally appearing in the arguments of $H_n$ to either the left (\ref{aln:PushingDerivativesLeft}) or the right (\ref{aln:PushingDerivativesRight}).
	\item Identify all contributions to the induced action up to and including quartic order $\cO(F^4)$.
	\item Evaluate the $y$-integrals present in $H_n$ using (\ref{eqn:H_nConstant}) and (\ref{eqn:H_nQuadratic}).
	\item Compute the outstanding momentum space integrals using Gaussian moments (\ref{eqn:a0GaussianMomentIdentity}).
	\item Express all resulting structures in terms of a chosen set of basis elements which are unique up to matrix trace and integration by parts.
\end{enumerate}

Omitting the details of the long calculations, we present here only the final result. 
We find that the leading contribution to $a_2$ for a generic NLED theory expanded in the weak-field regime is given by the following expression
\baa
	a_2=\frac{1}{\big(L_{\a}^{(0)}\big)^2}~\tr\bigg[&\frac{1}{12}\big(\pd^a\dsV_a^{(2)}\big)\big(\pd^b\dsV_b^{(2)}\big)+\frac{1}{24}\big(\pd^a\dsV^{b(2)}\big)\big(\pd_a\dsV_{b}^{(2)}\big)+\frac{1}{24}\big(\pd^a\dsV_a^{(2)}\big)\big(\Box\dsH^{b\ph{b}(2)}_{\ph{b}b}\big)\non\\
	&-\frac{1}{12}\big(\pd^a\dsV_a^{(2)}\big)\big(\pd^b\pd^c\dsH_{bc}^{(2)}\big)-\frac{1}{12}\big(\pd^a\dsV^{b(2)}\big)\big(\Box\dsH_{(ab)}^{(2)}\big)+\frac{1}{480}\big(\Box\dsH^{a\ph{a}(2)}_{\ph{a}a}\big)\big(\Box\dsH^{b\ph{b}(2)}_{\ph{b}b}\big)\non\\
	&+\frac{1}{240}\big(\Box\dsH^{ab(2)}\big)\big(\Box\dsH^{(2)}_{(ab)}\big)-\frac{1}{40}\big(\pd^a\pd^b\dsH_{ab}^{(2)}\big)\big(\Box\dsH^{c\ph{c}(2)}_{\ph{c}c}\big)\non\\
	&+\frac{1}{60}\big(\pd^a\pd^b\dsH^{(2)}_{ab}\big)\big(\pd^c\pd^d\dsH^{(2)}_{cd}\big)+\frac{1}{30}\big(\pd^a\pd_b\dsH^{(bc)(2)}\big)\big(\Box\dsH^{(2)}_{(ac)}\big)\bigg]+\cO(F^5)~.
	\label{e11}
\eaa

Once again, our result is consistent with other approaches.  For example, if we set $L_\a=-1$ and $\dsG_{ab}=0$ which corresponds to Maxwell theory, the operator \eqref{eqn:NLEDMatrixOperatorDefn} becomes minimal, and we recover the quartic terms generated by the Schwinger--DeWitt approach. Similarly, agreement with the results of~\cite{Grasso:2023qye} 
is obtained by solely setting $\dsG_{ab}=0$ and expanding up to and including quartic order.

Since $\dsH_{ab}^{(2)}$ and $\dsV_a^{(2)}$ are composite fields, we can further decompose our result in terms of $L_{\a}^{(2)}$, $L_{\b}^{(2)}$ and $G_{abcd}^{(2)}$, which yields
\baa \label{aln:NLEDa2GResult}
	a_2=\frac{1}{\big(L_{\a}^{(0)}\big)^2}\bigg[&\frac{1}{4}\big(\Box L_{\a}^{(2)}\big)\big(\Box L_{\a}^{(2)}\big)+\frac{1}{4}\big(\Box L_{\b}^{(2)}\big)\big(\Box L_{\b}^{(2)}\big)-\frac{1}{24}\big(\Box L_{\a}^{(2)}\big)\big(\Box G^{ab~(2)}_{\ph{ab}ba}\big)\non\\
	&+\frac{1}{12}\,\ve^{acde}\big(\Box L_{\b}^{(2)}\big)\big(\pd_a\pd^b G^{(2)}_{bcde}\big)+\frac{1}{480}\big(\Box G_{a\ph{b}bc}^{\ph{a}b(2)}\big)\big(\Box G^{cd\ph{d}a(2)}_{\ph{cd}d}\big)\non\\
	&+\frac{1}{240}\big(\Box G^{cabd(2)}\big)\big(\Box G_{d(ab)c}^{(2)}\big)+\frac{1}{60}\big(\pd^a\pd^bG^{(2)}_{cabd}\big)\big(\Box G^{de\ph{e}c(2)}_{\ph{de}e}\big)\non\\
	&+\frac{1}{60}\big(\pd^a\pd^bG^{(2)}_{cabd}\big)\big(\pd_e\pd_fG^{defc(2)}\big)-\frac{1}{40}\big(\pd^f\pd^bG_{cdbe}^{(2)}\big)\big(\pd_f\pd_aG^{eadc(2)}\big)\non\\
	&+\frac{1}{60}\big(\pd^f\pd^bG_{cdbe}^{(2)}\big)\big(\pd_f\pd_aG^{edac(2)}\big)\bigg]+\cO(F^5)~.
\eaa
Setting $G_{abcd}=0$, our result finds agreement with the quartic order expansion of Osborn~\cite{Osborn:2003vk} and the $n=1$ case of~\cite{Grasso:2023hmv}. Eqs.~\eqref{e11} and~\eqref{aln:NLEDa2GResult} represent the general result for the $a_2$ coefficient to leading order for an arbitrary NLED model. 

This result simplifies dramatically for the case of Born--Infeld theory, where the constant contributions of interest are
\be 
	L_{\a}^{(0)}=-1~,\qquad L_{\a\a}^{(0)}=L_{\b\b}^{(0)}=\frac{1}{T}~,\qquad L_{\b}^{(0)}=L_{\a\b}^{(0)}=0~,
\ee
and the quadratic contributions are given by 
\be 
	L_{\a}^{(2)}=\frac{\a}{T}~,\qquad L_{\b}^{(2)}=\frac{\b}{T}~.
\ee
Substituting this into~(\ref{aln:NLEDa2GResult}) yields
\baa \label{aln:QuarticA2BI}
	(a_2)_{\rm BI}=\frac{1}{T^2}\biggl[&-\frac{1}{60}\big(\Box \a\big)\big(\Box \a \big)+\frac{1}{4}\big(\Box \b\big)\big(\Box \b\big)-\frac{2}{15}\big(\Box\a\big)\pd^a\pd^b(F^2)_{ab}+\frac{1}{20}\Box(F^2)^{ab}\Box(F^2)_{ab}\non\\
	&-\frac{1}{10}\,\pd^d\pd^a(F^2)_{ab}\,\pd_d\pd_c(F^2)^{cb}+\frac{1}{30}\,\pd^a\pd^b(F^2)_{ab}\,\pd^c\pd^d(F^2)_{cd}\bigg]+\cO(F^5)~.
\eaa
We can simplify this result by finding a basis of Lorentz scalar, four derivative and quartic field strength structures using the procedure outlined in Appendix \ref{app:DerivativeBasis}. Our final result is
\baa
	(a_2)_{\rm BI}=\frac{1}{10T^2}\Big[(\pd_e&F_{ab})(\pd^eF^{bc})(\pd_f F_{cd})(\pd^fF^{da})+2(\pd_eF_{ab})(\pd_fF^{bc})(\pd^eF_{cd})(\pd^fF^{da})\Big]+\cO(F^5)~.
\eaa


\section{Heat kernel approach to conformal nonlinear electrodynamics}\label{sct:ConformalHeatKernelTechniques}

In this section, we adapt the Volterra-series approach from Subsection \ref{ssct:VolterraSeriesExpansion} to the subset of conformal NLED theories. With the exception of Maxwell electrodynamics, conformal NLED theories are non-analytic around zero and therefore excluded from the weak-field approximations of Subsections \ref{ssct:QuarticA0ContributionFlat}, \ref{ssct:QuarticA1ContributionFlat} and \ref{ssct:QuarticA2ContributionFlat}. While ModMax electrodynamics is the unique conformal and duality-invariant theory, there exists a wider class of purely conformal NLED theories. We recall that a conformal NLED theory satisfies the following property \eqref{eqn:ConformalNLEDProperty}
\be 
	\a L_{\a}+\b L_{\b}=L \qquad \implies \qquad L_{\a\a}L_{\b\b}-L_{\a\b}^2=0~.
\ee
Notably in this case, the background matrix $\dsG^{ab}$ (\ref{aln:dsGDefn}) inherits the following square property from its component tensor $G^{abcd}$ (see Appendix \ref{app:PropertiesofTensorG}):
\be \label{eqn:cG^2PropertyFlat}
	\dsG^{ab}\dsG^{cd}=G^{bc}\dsG^{ad}~,\qquad\qquad G^{ab}:=\tr\,\dsG^{ab}~.
\ee
For an arbitrary parameter $\t$, and introducing the notation $\cG(k):=-\ri k_ak_b \dsG^{ab}$, we find the exponential of $\cG(k)$ reduces to
\baa
	\re^{\t\, \cG(k)}&=\id_4+\Bigg(\sum_{n=0}^{\infty}\frac{\t^{n+1}}{(n+1)!}\big[\tr\, \cG(k)\big]^n\Bigg)\cG(k)\non\\
	&=\id_4+\bigg(\frac{\re^{\t\, \tr\,\cG(k)}-1}{\tr\,\cG(k)}\bigg)\cG(k)~.\label{aln:ExpcGPropertyFlat}
\eaa
This property is key to computing momentum space integrals that arise from the Volerra-series approach since it reduces the product of matrix structures to be at most linear in $\cG(k)$.

In Subsection \ref{ssct:A0ContributionFlat} we determine the $a_0$ contribution exactly,  showing that imposing causality conditions on the electromagnetic background is sufficient for convergence. In Subsection \ref{ssct:CausalRole}, we comment on the role of causality conditions on the convergence of the exact $a_1$ and $a_2$ contributions.


\subsection{The exact $a_0$ contribution}\label{ssct:A0ContributionFlat}

In this subsection, we study the $a_0$ contribution to the heat kernel of a conformal NLED theory, denoted $K(x;s)\big|_{a_0}$. This piece contains no derivatives on the background field strength and an integral of this form was studied for the case of ModMax electrodynamics in a constant background \cite{Pinelli:2021hkmm}. Following the Volterra-series approach \eqref{aln:VolterraSeriesResult}, the $a_0$ contribution is given by:
\be \label{eqnExactA0ContriFlat}
	K(x;s)\Big|_{a_0}=\frac{\tr}{s^2}\inte k\,\re^{-\ri k_ak_b \dsH^{ab}}=\frac{\tr}{s^2}\inte k\,\re^{\ri k^2 L_{\a}}\,\re^{\cG(k)}\,.
\ee
Applying identity \eqref{aln:ExpcGPropertyFlat} for conformal NLED to the $a_0$ contribution of the heat kernel (\ref{eqnExactA0ContriFlat}), we find
\be 
	K(x;s)\Big|_{a_0}=\frac{\tr}{s^2}\inte k\,\re^{\ri k^2 L_{\a}}\bigg[\id_4+\bigg(\frac{\re^{\tr\,\cG(k)}-1}{\tr\, \cG(k)}\bigg)\cG(k)\bigg]~.
\ee
Taking the trace simplifies the integral dramatically:
\baa \label{eqn:A0IntegralsFlat}
	K(x;s)\Big|_{a_0}&=\frac{3}{s^2}\inte k\,\re^{\ri k^2 L_{\a}}+\frac{1}{s^2}\inte k\,\re^{-\ri k_ak_b Z^{ab}}~ \non\\
	&=\frac{h}{s^2}\Bigg[\frac{3}{L_{\a}^2}+\frac{\re^{\frac{\ri\p}{4}(2-n_{+}+n_{-})}}{\sqrt{|\det(Z^{ab})|}}\Bigg]~,
\eaa
where
\be	\label{eqn:ZDefn}
	Z^{ab}:=-L_{\a}\h^{ab}+G^{ab}~,
\ee
and $n_{+}$ and $n_{-}$ denote the number of positive and negative eigenvalues of $Z^{ab}$, respectively. These results can be established by first diagonalising $Z^{ab}$, and then using the following Fresnel integral identities for real $a$ and $b$:
\be
	\int_{-\infty}^{\infty} \rd\t\, e^{\ri a\t^2+\ri b\t} = \left\{
	\begin{aligned}
		&\frac{1+\ri}{\sqrt{2}}\sqrt{\frac{\p}{a}}\exp\left(\frac{-\ri b^2}{4a}\right), \quad a>0\\
		&\frac{1-\ri}{\sqrt{2}}\sqrt{\frac{\p}{|a|}}\exp\left(\frac{\ri b^2}{4|a|}\right), \quad a<0
	\end{aligned}\right.~.
\ee
Comparing (\ref{eqn:A0IntegralsFlat}) and (\ref{eqn:DeWittCoefficientsFlat}), finally leads to the $a_0$ DeWitt coefficient
\be \label{eqn:NLEDA0FlatBefore}
	a_0=\frac{3}{L_{\a}^2}+\frac{\re^{\frac{\ri\p}{4}(2-n_{+}+n_{-})}}{\sqrt{|\det(Z^{ab})|}}~.
\ee

In Appendix \ref{app:PropertiesofTensorG} we demonstrate that the matrix $G:=(G^a_{\ph{a}b})$ has the property that its square can be expressed as
\be \label{eqn:G^2CharPoly}
	G^2+2A G-C\id_4=0~,\quad A:=-\frac{1}{4}\tr\,G~, \quad C:=\frac{1}{4}\bigg[\tr(G^2)-\hf (\tr\,G)^2\bigg]~,
\ee
Specifically, the quantities $A$ and $C$ are given by:
\baa
	A&=\a(L_{\a\a}-L_{\b\b})+2\b 		L_{\a\b}~,\label{eqn:G^2ADefnFlat}\\
	C&=4\a^2 L_{\a\a} L_{\b\b}+4\a\b L_{\a\b}(L_{\b\b}-L_{\a\a})+\b^2(L_{\a\a}+L_{\b\b})^2-4\b^2L_{\a\b}^2~.\label{eqn:G^2CDefnFlat}
\eaa
This implies that the composite, symmetric matrix $Z^{ab}$ has a determinant of the form\footnote{Appendix \ref{app:4x4MatrixProperties} includes full details.  In particular, the expression given here for $\det(Z^{ab})$ is analogous to \eqref{eqn:detZMatrix}.}
\be\label{eqn:DeterminantofZabFlat}
	\det(Z^{ab})=\det(-L_{\a}\h^{ab}+G^{ab})=-\big(L_{\a}^2+2A L_{\a} -C\big)^2~.
\ee
Thus the necessary and sufficient condition for convergence of the $a_0$ contribution is:
\be \label{eqn:Convergencea0Condition}
	a_0=\frac{3}{L_{\a}^2}+\frac{\re^{\frac{\ri\p}{4}(2-n_{+}+n_{-})}}{\sqrt{(L_{\a}^2+2A L_{\a} -C)^2}} \quad \Longleftrightarrow \quad L_{\a}^2+2A L_{\a} -C\neq0~.
\ee

We can simplify this result further by specialising to the subset of causal, conformal NLED theories. The two distinct eigenvalues of $Z=(Z^a_{\ph{a}b})=-L_{\a}\id_4+G$, each of algebraic multiplicity two, are:\footnote{In Appendix \ref{app:PropertiesofTensorG} we show that the weak-field causality conditions imply $C\geq0$.}
\baa \label{eqn:EivenvaluesofZFlat}
	\l_{\pm}&=-L_{\a}-A\pm\sqrt{A^2+B^2} \non\\
	&=-L_{\a}+\a(L_{\b\b}-L_{\a\a})-2\b L_{\a\b}\pm \big|L_{\a\a}+L_{\b\b}\big|\sqrt{\a^2+\b^2}~,
\eaa
having introduced $B:=\sqrt{C}$. It turns out that $\l_{-}>0$ is precisely the strong-field causality condition (\ref{eqn:BackgroundCausalCondition}), provided the weak-field causality conditions (\ref{eqn:BackgroundConvexCondition}) are also satisfied. This implies, in the case of causal NLED, all eigenvalues of $Z$ are positive definite $\l_{+}>\l_{-}>0$. In this case, we obtain $n_+=3$ and $n_-=1$.

Finally, we observe that the convergence condition (\ref{eqn:Convergencea0Condition}) can be written as follows
\be 
	L_{\a}^2+2A L_{\a} - B^2=\l_+\l_->0~.
\ee
As a result, the square root of the determinant for causal NLED is positive-definite
\be	\label{eqn:SignofsqrtZabFlat}
	\sqrt{|\det(Z^{ab})|}=\sqrt{(L_{\a}^2+2A L_{\a} -B^2)^2}=L_{\a}^2+2A L_{\a} -B^2>0~.
\ee
Therefore, causality of conformal NLED is sufficient for the convergence of the $a_0$ contribution
\be \label{eqn:NLEDA0Flat}
	a_0=\frac{3}{L_{\a}^2}+\frac{1}{L_{\a}^2+2A L_{\a} -B^2}~.
\ee
While not strictly necessary in this case, for the exact $a_1$ and $a_2$ contributions in Subsection \ref{ssct:CausalRole}, the strong-field causality condition on the background electromagnetic fields becomes necessary.

As an example, let us now apply the above result directly to a well-known causal, conformal NLED theory: ModMax electrodynamics. We obtain
\be 
	L_{\a}^2+2A L_{\a} -B^2=1~,
\ee
therefore
\be 
	(a_0)_{\rm MM}=\frac{3}{L_\a^2}+1=\frac{3}{\left(-\cosh\g+\frac{\a\sinh\g}{\sqrt{\a^2+\b^2}}\right)^2}+1~,
\ee
which agrees with Ref. \cite{Pinelli:2021hkmm}.


\subsection{On the role of causality in the exact $a_1$ and $a_2$ contributions}\label{ssct:CausalRole}

In this subsection, we study the relationship between causality and convergence of the heat kernel for conformal NLED theories. In this case, we give strong arguments that imposing the causality conditions on the background electromagnetic fields is necessary and sufficient for the convergence of both the exact $a_1$ and $a_2$ contributions. 

For computing the $a_1$ and $a_2$ contributions to all orders in the background electromagnetic field, the Volterra-series expansion generates a large number of distinct Gaussian tensor 
integrals, each with intricate matrix contractions and momentum space dependencies that must be evaluated systematically. 
Due to their computational complexity, we leave a complete treatment of this case for future work. Instead, we focus on the following integral
\be	
	\tr\inte k\,H_1[\dsQ]~,
\ee
where $\dsQ$ is an arbitrary matrix with no additional $k$-dependence. While terms of this type did not appear in the quartic order expansion, integrals of this form represent the simplest extension beyond the $a_0$ contribution into the exact $a_1$ and $a_2$ contributions. As such, evaluating this integral serves as a useful benchmark and offers insight into the causal structure of the heat kernel of a conformal NLED theory without a weak-field approximation. 

We begin by recalling the defnition (\ref{eqn:DefnofHIntegralsFlat})
\baa
	H_n[\dsQ_1\otimes \dsQ_2\otimes\cdots\otimes \dsQ_n]:=\re^\dsA&\int^1_0{\rm d}y_1\int^{y_1}_0{\rm d}y_2\cdots\int^{y_{n-1}}_0{\rm d}y_n\,\big(\re^{-y_1\dsA}\,\dsQ_1\,\re^{y_1\dsA}\big)\non\\
	&\qquad\times\big(\re^{-y_2\dsA}\,\dsQ_2\,\re^{y_2\dsA}\big)\times\cdots\times\big(\re^{-y_n\dsA}\,\dsQ_n\,\re^{y_n\dsA}\big)~.
\eaa
Compared to $n>1$, the $n=1$ case simplifies greatly due to the cyclic property of the trace
\be 
	\tr\,H_1[\dsQ]=\tr\bigg[\re^\dsA\int^1_0{\rm d}y\,\big(\re^{-y\dsA}\,\dsQ\,\re^{y\dsA}\big)\bigg]=\tr\big[\re^{-\ri k_ak_b \dsH^{ab}}\,\dsQ\big]~.
\ee
What remains is to compute the outstanding $k$-integral
\baa 
	\tr\inte k\,H_1[\dsQ]&=\tr\inte k\,\re^{\ri k^2L_{\a}}\big[\re^{\cG(k)}\dsQ\big]~ \non\\
	&=\tr\inte k\,\re^{\ri k^2 L_{\a}}\bigg[\dsQ+\bigg(\frac{\re^{\tr\,\cG(k)}-1}{\tr\,\cG(k)}\bigg)\cG(k)\dsQ\bigg]\label{eqn:OriginalH1BInt}~,
\eaa
where here $\cG(k):= -\ri k_ak_b\dsG^{ab}$, and we have used the exponential identity of $\cG(k)$ (\ref{aln:ExpcGPropertyFlat}). The first term is routine,
\be \label{eqn:H1EasyInt}
	\tr(\dsQ)\inte k\,\re^{\ri k^2 L_{\a}}=\frac{h}{L_{\a}^2}\,\tr(\dsQ)~,
\ee
but the second term presents a more subtle challenge. To evaluate it, we adopt a Feynman-like integral trick.
The fundamental idea is to extend our expressions to a family of parametrised $\t$-integrals, which encompass the original integrals. 
The parameter $\t$ is chosen such that differentiating with respect to it yields a more tractable momentum space integral.  We then recover a result for the original integral by integrating over $\t$.

To evaluate the second term in (\ref{eqn:OriginalH1BInt}), consider the following $\t$-parametrised family of integrals
\be
	I(\t):=\inte k\,\re^{\ri k^2 L_{\a}}\bigg(\frac{\re^{\t\,\tr\,\cG(k)}-1}{\tr\,\cG(k)}\bigg)\tr(\cG(k) \dsQ)~.
\ee
We can recover the original integral by setting $\t=1$, and setting $\t=0$ leads to the notable result $I(0)=0$. Differentiatating with respect to $\t$ gives
\be 
	I'(\t)=\inte k\,\re^{\ri k^2 L_{\a}+\t\,\tr\,\cG(k)}\,\tr(\cG(k) \dsQ)~,
\ee
which reduces to a standard Gaussian integral. Extracting the $k$-dependence from $\cG(k)$ and combining the above results together, we find the original expression we wish to solve is given by
\be \label{eqn:OriginalQuantityFeynman}
	I(1)=\int^1_0\rd \t\,I'(\t)=-\ri\,\tr(\dsG^{ab}\dsQ)\int^1_0\rd \t\inte k\,\re^{-\ri k_ck_dZ^{cd}(\t)}k_ak_b~,
\ee
where we have introduced the matrix
\be\label{eqn:ZTauDefn}
	Z^{ab}(\t):=-L_{\a}\h^{ab}+\t G^{ab}~,
\ee 
which is a $\t$-parametrisation of an earlier introduced $Z^{ab}(1)=Z^{ab}$ (\ref{eqn:ZDefn}). Evaluating the $k$-integral yields
\be\label{eqn:IntegrandBeforeTauIntegral}
	\inte k\,\re^{-\ri k_ck_dZ^{cd}(\t)}k_ak_b=\biggl(-\frac{\ri h}{2}\biggr)\frac{\re^{\frac{\ri\p}{4}(2-n_{+}(\t)+n_{-}(\t))}}{\sqrt{|\det{(Z^{cd}(\t))}|}}\,(Z^{-1})_{ab}(\t)~,
\ee
where $n_{+}(\t)$, $n_{-}(\t)$ denotes the number of positive and negative eigenvalues of $Z^{cd}(\t)$, respectively.

As in Subsection \ref{ssct:A0ContributionFlat}, we highlight that the matrix $G=(G^a_{\ph{a}b})$ has the property that its square can be expressed as
\be\label{eqn:CommentSquareGProperty}
	G^2+2A G-C\id_4=0~,\quad A:=-\tfrac{1}{4}\tr\,G~, \quad C:=\tfrac{1}{4}\big[\tr(G^2)-\tfrac{1}{2} (\tr\,G)^2\big]~,
\ee
where:
\baa
	A&=\a(L_{\a\a}-L_{\b\b})+2\b L_{\a\b}~,\\
	C&=4\a^2 L_{\a\a} L_{\b\b}+4\a\b L_{\a\b}(L_{\b\b}-L_{\a\a})+\b^2(L_{\a\a}+L_{\b\b})^2-4\b^2L_{\a\b}^2~.
\eaa
This property, in conjuction with results from Appendix \ref{app:4x4MatrixProperties}, implies the determinant of $Z^{ab}(\t)$ is of the form
\be
	\det(Z^{ab}(\t))=\det(-L_{\a}\h^{ab}+\t G^{ab})=-\big(L_{\a}^2+2A L_{\a}\t -C\t^2\big)^2~,
\ee
and the inverse matrix, $(Z^{-1})_{ab}(\t)$, of the form
\be 
	(Z^{-1})_{ab}(\t)=-\bigg(\frac{L_{\a}+2A\t}{L_{\a}^2+2A L_{\a}\t-C\t^2}\bigg)\,\h_{ab}- \bigg(\frac{\t}{L_{\a}^2+2A L_{\a}\t-C\t^2}\bigg)\,G_{ab}~.
\ee
The two distinct eigenvalues of $Z(\t)=(Z^a_{\ph{a}b}(\t))=-L_{\a}\id_4+\t G$, each with algebraic multiplicity two, are:
\baa \label{eqn:EigenvaluesofZtau}
	\l_{\pm}(\t)&=-L_{\a}-A\t\pm\sqrt{(A^2+C)\t^2}\non\\
	&=-L_{\a}+\big[\a(L_{\b\b}-L_{\a\a})-2\b L_{\a\b}\big]\t\pm \big|L_{\a\a}+L_{\b\b}\big|\sqrt{(\a^2+\b^2)\t^2}~.
\eaa
At $\t=0$, both eigenvalues reduce to $\l_{\pm}(0)=-L_{\a}>0$, which is the background sign convention (\ref{eqn:BackgroundSignConvention}). At $\t=1$, either $\l_+(1)$ or $\l_-(1)$ corresponds to the strong-field causality condition (\ref{eqn:BackgroundCausalCondition}), depending on the sign of $L_{\a\a}+L_{\b\b}$.

We now clarify why the strong-field causality condition is necessary for convergence of the $a_1$ and $a_2$ contribution. Since the integrand (\ref{eqn:IntegrandBeforeTauIntegral}) involves both the determinant and inverse matrix of $Z(\t)$, a condition for convergence of these integrals is a $\t$-parameter extension of the earlier condition for convergence required for the exact $a_0$ contribution (\ref{eqn:Convergencea0Condition}):
\be 
	L_{\a}^2+2A L_{\a}\t -C\t^2=\l_+(\t)\l_-(\t)\neq0~,\qquad \t\in[0,1]\,.
\ee 
Without loss of generality, let us assume of the two eigenvalues, $\l_-(1)\leq0$ is the violated strong-field causality condition (\ref{eqn:BackgroundCausalCondition}). Combined with $\l_-(0)>0$ from (\ref{eqn:BackgroundSignConvention}) and the fact $\l_-(\t)$ is a linear function in $\t\in[0,1]$, it follows there exists $\t_0\in(0,1]$ such that $\l_-(\t_0)=0$. This implies that violation of the strong-field 
causality condition leads to a divergent $\t$-integral in the region $\t\in[0,1]$.

We now demonstrate that both the weak and strong-field causality conditions are sufficient for the convergence of the $a_1$ and $a_2$ contributions. 
From here onward we will restrict our attention to only causal, conformal NLED models. In causal NLED, all eigenvalues of $Z(\t)$ 
are positive definite $\l_{+}(\t)\geq\l_{-}(\t)>0$ for $\t\in[0,1]$, since $\l_-(\t)$ is directly related to the strong-field causality condition. Consequently, the matrix $Z^{cd}(\t)$ 
has a fixed signature with $n_+(\t)=3$ and $n_{-}(\t)=1$ for $\t\in[0,1]$. We also observe that within the region $\t\in[0,1]$ 
the quantity\footnote{In Appendix \ref{app:PropertiesofTensorG}, we show that the weak-field causality conditions imply $C\geq0$.}
\be 
	L_{\a}^2+2A L_{\a}\t -B^2\t^2=\l_+(\t)\l_-(\t)>0~,\qquad B:=\sqrt{C}
\ee
remains strictly positive. Consequently, the square root of the determinant has well-defined sign within the region
\be 
	\sqrt{|\det(Z^{ab}(\t))|}=\sqrt{(L_{\a}^2+2A L_{\a}\t -B^2\t^2)^2}=L_{\a}^2+2A L_{\a}\t -B^2\t^2>0~.
\ee

Combining the above results into the original integral (\ref{eqn:OriginalQuantityFeynman}) that we wish to evaluate 
\baa \label{eqn:PostEvaluatingkIntegral}
	I(1)&=-\frac{h}{2}\,\tr(\dsG^{ab}\dsQ)\int^1_0\rd \t\,\frac{\re^{\frac{\ri\p}{4}(2-n_+(\t)+n_-(\t))}}{\sqrt{|\det{(Z^{cd}(\t))}|}}\,(Z^{-1})_{ab}(\t)\non\\
	&=\frac{h}{2}\,\tr(\dsG^{ab}\dsQ)\bigg\{\h_{ab}\int^1_0\rd \t\,\frac{L_{\a}+2A\t}{(L_{\a}^2+2A L_{\a}\t-B^2\t^2)^2}\non\\
	&\qquad\qquad\qquad\qquad\quad +G_{ab}\int^1_0\rd \t\,\frac{\t}{(L_{\a}^2+2A L_{\a}\t-B^2\t^2)^2}\bigg\}~.
\eaa
We can evaluate the integrals (\ref{eqn:PostEvaluatingkIntegral}) by noting the general result for $\a_1,\a_2\in\dsR$ and for $C_0$ an integration constant:
\baa
	&\int\rd \t\,\frac{\a_1+\a_2\,\t}{(L_\a^2+2A L_{\a} \t-B^2\t^2)^2}\non\\
	&\non\\[-1em]
	&=\left\{
	\begin{aligned}
		&\frac{\t(2\a_1+\a_2\t)}{L_{\a}^4}+C_0~,\qquad A=B=0~,\\
		&\\[-1.5em]
		& \frac{-2\a_1A+\a_2L_{\a}}{4A^2L_{\a}(L_\a^2+2A L_{\a} \t)}+\frac{\a_2}{4A^2L_{\a}^2}\ln(L_{\a}^2+2A L_{\a}\t)+C_0~,\quad 	A\neq0,~B=0~,\\
		&\\[-1.5em]
		& \frac{\a_1(-A L_{\a}+B^2\t)+\a_2(L_{\a}^2+A L_{\a}\t)}{2L_{\a}^2(A^2+B^2)(L_{\a}^2+2A L_{\a}\t-B^2\t^2)}\\
		&\quad+\frac{\a_1B^2+\a_2A L_{\a}}{2L_{\a}^3(A^2+B^2)^\frac{3}{2}}\,\arctanh\bigg(\frac{-A L_{\a}+B^2\t}{L_{\a}\sqrt{A^2+B^2}}\bigg)+C_0~,\quad B\neq0~.
	\end{aligned}
	\right.
\eaa
As a consequence of the causality conditions (\ref{eqn:BackgroundConvexCondition}) and (\ref{eqn:BackgroundCausalCondition}), the logarithm and $\arctanh$ functions prove to be well-defined in the region $\t\in[0,1]$. To evaluate the integrals in expression (\ref{eqn:PostEvaluatingkIntegral}), we split into three distinct cases:
\paragraph{\underline{$A=B=0$}}
\be
	I(1)=\frac{h}{4 L_{\a}^4}\,\tr(\dsG^{ab}\dsQ)\Big\{2\h_{ab}L_{\a}+G_{ab}\Big\}~.
\ee
\paragraph{\underline{$A\neq0,~B=0$}}
\be
	I(1)=\frac{h}{4A L_{\a}^2}\,\tr(\dsG^{ab}\dsQ)\bigg\{\bigg[\h_{ab}+\frac{G_{ab}}{2A}\bigg]\ln\bigg(1+\frac{2A}{L_{\a}}\bigg)-\frac{G_{ab}}{L_{\a}+2A}\bigg\}~.
\ee
\paragraph{\underline{$B\neq0$}}
\baa
	I(1)&=\frac{h}{4(A^2+B^2)L_{\a}^2}\,\tr(\dsG^{ab}\dsQ)\bigg\{\frac{\h_{ab}\,B^2(L_{\a}+A)+G_{ab}(B^2-A L_{\a})}{(L_{\a}^2+2A L_{\a}-B^2)}\non\\
	&\qquad\qquad\qquad\qquad\qquad\qquad+\bigg[\frac{\h_{ab}(B^2+2A^2)+G_{ab}A}{\sqrt{A^2+B^2}}\bigg]\arctanh\bigg(\frac{\sqrt{A^2+B^2}}{L_{\a}+A}\bigg)\bigg\}~.
\eaa
We can simplify further by noting the property of the matrix $G=(G^a_{\ph{a}b})$ (\ref{eqn:CommentSquareGProperty})
\be
	\dsG^{ab}\h_{ab}=G~,\qquad \dsG^{ab}G_{ab}=G^2=-2A G+B^2\id_4~,
\ee
which consolidates all cases into one unified form
\baa
	I(1)&=\frac{h}{4}\Bigg\{\tr(\dsQ)\bigg[\frac{1}{L_{\a}^2+2A L_{\a}-B^2}-\frac{1}{L_{\a}^2}\bigg]\non\\
	&\quad+\frac{\tr\big([A\id_4+G]\dsQ\big)}{(A^2+B^2)L_{\a}^2}\bigg[\frac{B^2(L_{\a}-A)+2A^2 L_{\a}}{L_{\a}^2+2A L_{\a}-B^2}+\frac{B^2}{\sqrt{A^2+B^2}}\,\arctanh\bigg(\frac{\sqrt{A^2+B^2}}{L_{\a}+A}\bigg)\bigg]\Bigg\}~,
\eaa
and finishes the calculation for the second term in (\ref{eqn:OriginalH1BInt}). Combined with the result for the first term (\ref{eqn:H1EasyInt}), we find expression (\ref{eqn:OriginalH1BInt}) is given by
\baa \label{aln:H1QIntResult}
	&\tr\inte k\,H_1[\dsQ]\non\\
	&=\frac{h}{4}\Bigg\{\tr(\dsQ)\bigg[\frac{1}{L_{\a}^2+2A L_{\a}-B^2}+\frac{3}{L_{\a}^2}\bigg]\non\\
	&\qquad+\frac{\tr\big([A\id_4+G]\dsQ\big)}{(A^2+B^2)L_{\a}^2}\bigg[\frac{B^2(L_{\a}-A)+2A^2 L_{\a}}{L_{\a}^2+2A L_{\a}-B^2}+\frac{B^2}{\sqrt{A^2+B^2}}\,\arctanh\bigg(\frac{\sqrt{A^2+B^2}}{L_{\a}+A}\bigg)\bigg]\Bigg\}~.
\eaa
Having obtained this result we have demonstrated that both causality conditions are sufficient for convergence.
As a consistency check for the above result we can set $\dsQ=\id_4$
\be 
	\tr\inte k\,H_1[\id_4]=\tr\inte k\re^{-\ri k_ak_b\dsH^{ab}}=K(x;s)\Big|_{a_0}~,
\ee
which was calculated in Subsection \ref{ssct:A0ContributionFlat}. We indeed recover the previously calculated $a_0$ result in the case of causal, conformal NLED (\ref{eqn:NLEDA0Flat}) from (\ref{aln:H1QIntResult}), since we find from the definition of $A$ (\ref{eqn:CommentSquareGProperty})
\be
	\tr\big(A\id_4+G\big)=0~.
\ee

As an example, let us apply the general expression (\ref{aln:H1QIntResult}) directly to a causal, conformal NLED theory: ModMax electrodynamics. A useful simplification of the $\arctanh$ function occurs immediately for ModMax theory, since
\be 
	\arctanh\bigg(\frac{\sqrt{A^2+B^2}}{L_{\a}+A}\bigg)=-\arctanh(\tanh\g)=-\g~,
\ee
which we use to obtain
\baa 
	\bigg(\tr\inte k\,H_1[\dsQ]\bigg)_{\rm MM}&=\frac{h}{4}\Bigg\{\tr(\dsQ)\bigg[\frac{3}{(-\cosh\g+\frac{\a\sinh\g}{\sqrt{\a^2+\b^2}})^2}+1\bigg]\non\\
	&\qquad\quad-\frac{\tr\big([\a\id_4+F^2]\dsQ\big)}{(-\cosh\g+\frac{\a\sinh\g}{\sqrt{\a^2+\b^2}})^2(\a^2+\b^2)^{3/2}}\non\\
	&\qquad\qquad\times\Big[(2\a^2+\b^2)\cosh\g\sinh\g-2\a\sqrt{\a^2+\b^2}\sinh^2\g+\b^2\g\Big]\Bigg\}~.
\eaa


\section{Conclusion}\label{sct:Outlook}

While much attention has been devoted to the classical structure of nonlinear electrodynamics (NLED), the question of its consistent quantisation remains open and of great interest. 
Whether viewed as effective actions descending from higher-dimensional reductions or as standalone models, the quantum behaviour of these theories continues to present subtle challenges. 
In this paper we developed a systematic approach to study one-loop effective actions for non-minimal matrix operators which naturally arise in this context.
Our approach used the Volterra-series formalism considered previously in Refs.~\cite{Iochum:2016ynh, Iochum:2017ver, Grasso:2023qye}.
We computed the quartic contributions up to the $a_0$, $a_1$ and $a_2$ DeWitt coefficients for a general NLED model analytic around zero and obtained the exact $a_0$ coefficient in the case of conformal NLED.

The structure of the induced action arising from the $a_2$ coefficient remains only partially understood. For NLED models with conformal and/or duality invariance, one expects these 
symmetries to be inherited by the induced action (see \cite{Fradkin:1984qedf,Roiban:2012dspqt} for formal arguments). 
Using these assumptions, the ModMax induced action has been determined  up to overall constants in Ref.~\cite{Kuzenko:2024hdmm}. 
Preliminary results obtained in Subsection \ref{ssct:CausalRole} suggest that for a conformal NLED model, causality is both necessary and sufficient for convergence of the induced action.
It would be of some interest to extend the heat kernel methods developed in this paper to obtain its complete structure.

From the derivative expansion developed in Subsection \ref{ssct:VolterraSeriesExpansion}, we find that for a constant background field strength the $a_1$ and $a_2$ coefficients vanish identically.
In curved backgrounds, however, these coefficients may receive non-zero contributions, including purely gravitational curvature terms and mixed curvature and field strength products. 
To evaluate these contributions, one may employ background field techniques analogous to those used beyond the one-loop level \cite{Kuzenko:2003bfmop},
involving the Synge--DeWitt world  function and the parallel displacement propagator. 
In the coincidence limit, the covariant derivatives of these structures admit known curvature expansions. 
Additionally, when the background field strength is covariantly constant, the curvature itself may be expressed in terms of the field strength. 
This would allow the resulting induced action to be understood as a new NLED theory on a curved background.


\section*{Acknowledgements}

The authors would like to thank Sergei Kuzenko for collaboration at early stages of the project and for useful discussions and suggestions. 
The work of E.I.B is supported in part by the Australian Research Council, 
project DP230101629. The work of J.R.P is supported by the Australian Government Research Training Program Scholarship. We acknowledge the use of Wolfram Mathematica in carrying out some laborious computations associated with the heat kernel coefficients.


\appendix

\section{Matrix properties in four dimensions}\label{app:4x4MatrixProperties}

In this appendix we collect a set of matrix identities central to the calculations in the main text.  We first recall that in four dimensions an arbitrary $4\times4$ matrix $M=(M^a_{\ph{a}b})$ has a determinant which may be expressed as
\be\label{eqn:4x4Det}
	\det M=\tfrac{1}{4!}\Big[(\tr M)^4-6(\tr M)^2\,\tr (M^2)+8(\tr M)\tr (M^3)+3\big(\tr (M^2)\big)^2-6\,\tr (M^4)\Big]~,
\ee
and, from Cayley--Hamilton's theorem, will satisfy 
\baa\label{eqn:4x4CharMatrixEqn}
	0=M^4-&(\tr M)M^3+\tfrac{1}{2}\big[(\tr M)^2-\tr (M^2)\big]M^2\non\\
	&-\tfrac{1}{3!}\big[(\tr M)^3-3(\tr M)\tr (M^2)+2\,\tr (M^3)\big]M+(\det M)\,\id_4~.
\eaa
In the case of the electromagnetic field strength $F_{ab}=-F_{ba}$, the expression (\ref{eqn:4x4CharMatrixEqn}) for the traceless matrix $F=(F^a_{\ph{a}b})$ reduces to 
\be \label{eqn:AntisymmetricCharPol}
	0=F^4+2\a F^2-\b^2\id_4~,
\ee
where 
\be
	\a:=-\tfrac{1}{4}\tr(F^2)~,\qquad~
	\b^2:=\tfrac{1}{4}\big[\tr(F^4)-\tfrac{1}{2} (\tr(F^2))^2\big]=-\det F\geq 0~,
\ee
are the electromagnetic invariants defined in \eqref{NLEDLagrangian}.  The general identity (\ref{eqn:AntisymmetricCharPol}) then carries several implications for the following composite matrices  with constants $p,q\in\dsR$:
\be\label{eqn:4x4CompositeMatrix}
\left\{
\begin{array}{l}
	P:=p\id_4+q F~,\\
	Q:=p\id_4+q F^2~.
\end{array}
\right.
\ee

For the first, the determinant takes the form
\be
	\det P=\det\big(p\id_4+qF\big)=p^4+2\a p^2q^2-\b^2q^4~,
\ee
with four eigenvalues 
\be 
	\l_{\pm\pm}=p\pm\sqrt{(-\a\pm\sqrt{\a^2+\b^2})q^2}~,
\ee
one of the consequences of which is the well-known fact that the Born--Infeld Lagrangian (\ref{eqn:BornInfeldLagrangian}) may be expressed as
\be
	L_{\rm BI}(\a,\b)=T-T\sqrt{-\det\big(\h_{ab}+T^{-\hf}F_{ab}\big)}=T-\sqrt{T^2+2\a T-\b^2}~.
\ee

The symmetric matrix $Q$ appears frequently in the context of NLED, particularly in the composite matrices $G$ \eqref{eqn:trGForm} and $Z$ \eqref{eqn:ZDefn}, \eqref{eqn:ZTauDefn}. Its determinant takes the form
\be\label{eqn:detZMatrix}
	\det Q=\det\big(p\id_4+q F^2\big)=\big(p^2-2\a pq -\b^2q^2\big)^2~,
\ee
with two distinct eigenvalues each of algebraic multiplicity 2:
\be
	\l_{\pm}=p-\a q \pm \sqrt{(\a^2+\b^2)q^2}~.
\ee
As a consequence of (\ref{eqn:AntisymmetricCharPol}), $Q$ satisfies the quadratic identity 
\be
	Q^2+2A Q-C\id_4=0~,\qquad A:=-\tfrac{1}{4}\tr\,Q~, \qquad C:=\tfrac{1}{4}\big[\tr(Q^2)-\tfrac{1}{2} (\tr\,Q)^2\big]~.
\ee
Specifically, the quantities $A$ and $C$ are given in terms of $\a$ and $\b$ by:
\begin{subequations} \label{subeqn:TildeConstants}
	\baa 
		A&=-p+\a q~,\\
		C&=-p^2+2\a pq+\b^2q^2~.
	\eaa
\end{subequations}
This leads to the inverse matrix $Q^{-1}$ taking a simple form
\be
Q^{-1}=(p\id_4+q F^2)^{-1}=\bigg[\frac{p-2\a q}{p^2-2\a pq-\b^2q^2}\bigg]\id_4- \biggl[\frac{q}{p^2-2\a pq-\b^2q^2}\biggr]F^2~.
\ee


\section{Basis for Lorentz scalar, derivative and field strength structures}\label{app:DerivativeBasis}

In this appendix, we aim to classify all Lorentz scalar, four derivative, quartic field strength structures which appear in the quartic order $a_2$ contributions for Born--Infeld theory (\ref{aln:QuarticA2BI})
\baa
	(a_2)_{\rm BI}=\frac{1}{T^2}\biggl[&-\frac{1}{60}\big(\Box \a\big)\big(\Box \a \big)+\frac{1}{4}\big(\Box \b\big)\big(\Box \b\big)-\frac{2}{15}\big(\Box\a\big)\pd^a\pd^b(F^2)_{ab}+\frac{1}{20}\Box(F^2)^{ab}\Box(F^2)_{ab}\non\\
	&-\frac{1}{10}\,\pd^d\pd^a(F^2)_{ab}\,\pd_d\pd_c(F^2)^{cb}+\frac{1}{30}\,\pd^a\pd^b(F^2)_{ab}\,\pd^c\pd^d(F^2)_{cd}\bigg]+\cO(F^5)~.
	\label{B1}
\eaa
Similar quartic order corrections have been found in effective actions for open and type II string theories \cite{Andreev:1988cb, Leigh:1989jq, Shmakova:1999ai, Wyllard:2000qe}, and other supersymmetric contexts 
(see e.g. \cite{DeGiovanni:1999hr, Tseytlin:1999dj, Koerber:2002zb, Chemissany:2006qd, Grasso:2007zfa}). These structures were arranged into a basis for four derivative, quartic structures modulo the lowest order equations of motion, the Bianchi identity and integration by parts
\be \label{eqn:QuarticString}
	\frac{\k}{T^2}\bigg(\frac{1}{8}X_1+\frac{1}{4}X_2-\frac{1}{2}X_4-X_3\bigg)~.
\ee
Here $\k$ is some multiplicative constant and the basis is defined as folllows:
\begin{subequations}\label{aln:XBasisB}
	\baa 
	X_1&:=(\pd_eF_{ab})(\pd^eF^{ba})(\pd_fF_{cd})(\pd^fF^{dc})~,\\
	X_2&:=(\pd_eF_{ab})(\pd_fF^{ba})(\pd^eF_{cd})(\pd^fF^{dc})~,\\
	X_3&:=(\pd_eF_{ab})(\pd^eF^{bc})(\pd_f F_{cd})(\pd^fF^{da})~,\\
	X_4&:=(\pd_eF_{ab})(\pd_fF^{bc})(\pd^eF_{cd})(\pd^fF^{da})~.
	\eaa
\end{subequations}

Let us rewrite our result~\eqref{B1} in terms of these basis structures. 
At the level of Maxwell electrodynamics, the field strength $F_{ab}$ satisfies two fundamental relations, the equations of motion and the Bianchi identity, respectively;
\be 
\pd_aF^{ab}=0~, \qquad\qquad \pd_{[a}F_{bc]}=0~.
\ee
When extending to a general NLED model $L(\a,\b)$ only the equations of motion are altered
\be
L_{\a}(\pd_aF^{ab})+\hf G^{abcd}(\pd_aF_{cd})=0~,
\ee
where the tensor field $G^{abcd}$ contains second derivatives of the Lagrangian
\be
G^{abcd}:=\big(L_{\a\a}F^{ab}+L_{\a\b}\~F^{ab}\big)F^{cd}+\big(L_{\a\b}F^{ab}+L_{\b\b}\~F^{ab}\big)\~F^{cd}~.
\ee
When considering a weak-field regime, the leading contribution to the equations of motion coincides with Maxwell's up to cubic corrections
\be 
\pd_aF^{ab}=0+\cO(F^3)~.
\ee
Together with the Bianchi identity, this implies
\be 
\Box F^{ab}=0+\cO(F^3)~.
\ee
With these results, it follows that we can relate the following three types of four-derivative, quartic $F_{ab}$ structures:\footnote{Here we are using a schematic notation, the various indices and the exact patterns of contraction has been suppressed.}
\be
\left\{
\begin{array}{l}
	FF(\pd^2F)(\pd^2F)~,\\
	F(\pd F)(\pd F)(\pd^2F)~,\\
	(\pd F)^4~.
\end{array}
\right.\non
\ee 
Let us show that every element of the form $FF(\pd^2F)(\pd^2F)$ is related to $F(\pd F)(\pd F)(\pd^2F)$ by integration by parts. Imposing the equations of motion and Bianchi identity, only three types of terms are non-zero, modulo $\cO(F^5)$ contributions. In a schematic notation (some indices have been suppressed), integrating by parts yields:
\begin{subequations}
	\baa
		FF(\pd^a\pd F)(\pd_a\pd F)=-(\pd^a F)F(\pd F)(\pd_a \pd F)-F(\pd^a F)(\pd F)(\pd_a \pd F)&+\cO(F^5)~,\\
		FF(\pd_a\pd F)(\pd\pd F^{ab})=-(\pd^a F)F(\pd F)(\pd \pd F^{ab})-F(\pd^a F)(\pd F)(\pd \pd F^{ab})&+\cO(F^5)~,\\
		F_{ac}F_{db}(\pd^a\pd^b F^{ef})(\pd^c\pd^d F_{ef})=-F_{ac}(\pd^a F_{db})(\pd^b F^{ef})(\pd^c\pd^dF_{ef})&+\cO(F^5)~.
	\eaa
\end{subequations}
Next let us consider elements of the form $F(\pd F)(\pd F)(\pd^2F)$. In particular, let us focus on possible contractions of $\pd^2$ with the remaining stuctures. Imposing the equations of motion and Bianchi identity, only two types of terms are non-zero, modulo $\cO(F^5)$ contributions. Either one derivative is contracted with the bare $F$, or both derivatives are fully contracted with the $(\pd F)(\pd F)$ factor. The former proves to be related to the latter using the Bianchi identity. Any element $F(\pd F)(\pd F)(\pd^2F)$ of the latter form, can be related to the following thirteen terms by the Bianchi identity:
\medskip
\begin{minipage}{0.45\textwidth}
	\bna
		W_1&:=F_{ab}(\pd^e F^{ac})(\pd^f F^{bd})(\pd_e\pd_f F_{cd})~,\\
		W_2&:=F_{ab}(\pd^e F^{ac})(\pd^d F^{fb})(\pd_e\pd_f F_{cd})~,\\
		W_3&:=F_{ab}(\pd^c F^{ea})(\pd^d F^{fb})(\pd_e\pd_f F_{cd})~,\\
		W_4&:=F_{ab}(\pd^e F^{fa})(\pd^c F^{bd})(\pd_e\pd_f F_{cd})~,\\
		W_5&:=F_{ab}(\pd^c F^{ea})(\pd_d F^f_{\ph{f}c})(\pd_e\pd_f F^{bd})~,\\
		W_6&:=F_{ab}(\pd^e F^{ac})(\pd^f F_{cd})(\pd_e\pd_f F^{bd})~,\\
		W_7&:=F_{ab}(\pd^e F^{ac})(\pd^f F_{cd})(\pd_e\pd_f F^{bd})~,
	\ena
\end{minipage}%
\begin{minipage}{0.45\textwidth}
	\bna
		W_8&:=F_{ab}(\pd^e F^{fd})(\pd_d F^a_{\ph{a}c})(\pd_e\pd_f F^{bc})~,\\
		W_9&:=F_{ab}(\pd^e F^{fd})(\pd_c F^a_{\ph{a}d})(\pd_e\pd_f F^{bc})~,\\
		W_{10}&:=F_{ab}(\pd^e F^{cd})(\pd_c F^{fa})(\pd_e\pd_f F^b_{\ph{b}d})~,\\
		W_{11}&:=F_{ab}(\pd^c F^{ed})(\pd_c F^f_{\ph{f}d})(\pd_e\pd_f F^{ab})~,\\
		W_{12}&:=F_{ab}(\pd^c F^{ed})(\pd_d F^f_{\ph{f}c})(\pd_e\pd_f F^{ab})~,\\
		W_{13}&:=F_{ab}(\pd^e F^{fc})(\pd^d F^{ab})(\pd_e\pd_f F_{cd})~.\\
	\ena
\end{minipage}

\medskip
\noindent
Using integration by parts, the $W_i$ basis is related to the $X_i$ structures listed in (\ref{aln:XBasisB}) as  follows

\vspace{-1em}
\begin{minipage}{0.45\textwidth}
	\bna
	W_1&=-X_3+\hf X_4~,\\
	W_2&=\frac{1}{32}(X_1-6X_2+24X_3-12X_4)~,\\
	W_3&=\frac{1}{16}(-X_1+6X_2-8X_3+4X_4)~,\\
	W_4&=\frac{1}{32}(-X_1+6X_2+8X_3-4X_4)~,\\
	W_5&=0~,\\
	W_6&=-\hf X_4~,\\
	W_7&=\frac{1}{32}(-3X_1+10X_2-8X_3+20X_4)~,
	\ena
\end{minipage}%
\begin{minipage}{0.45\textwidth}
	\bna
	W_8&=0~,\\
	W_9&=\frac{1}{32}(-X_1-2X_2+8X_3+12X_4)~,\\
	W_{10}&=\frac{1}{32}(X_1+2X_2-8X_3+4X_4)~,\\
	W_{11}&=\frac{1}{8}(X_1+2X_2-8X_3+4X_4)~,\\
	W_{12}&=\frac{1}{8}(-X_1+6X_2-8X_3+4X_4)~,\\
	W_{13}&=\frac{1}{4}X_1~.\\
	\ena
\end{minipage}

\medskip
\noindent
Using this reduction procedure, our result from the heat kernel approach~\eqref{B1} can now be written in the form
\be 
	(a_2)_{\rm BI}=\frac{1}{T^2}\bigg(\frac{1}{80}X_1-\frac{1}{8}X_2+\frac{1}{5}X_3+\frac{1}{4}X_4\bigg)+\cO(F^5)~.
\ee
It can be shown that in four dimensions the structures~\eqref{aln:XBasisB} are further linearly dependent,
which is a consequence of the vanishing of antisymmetrisation in five indices:
\be 
	(\pd_eF^{[ab})(\pd^eF_{bc})(\pd_f F^{cd]})(\pd^fF^{da})=0~,\qquad (\pd^eF^{[ab})(\pd^fF_{bc})(\pd_e F^{cd]})(\pd_fF^{da})=0~.
\ee
This allows us to further reduce the basis using:
\be
	X_1=\frac{4}{3}(4X_3-3X_4)+\cO(F^5)~,\qquad X_2=\frac{4}{3}X_3+\cO(F^5)~,
\ee
and express our final result as
\be 
	(a_2)_{\rm BI}=\frac{1}{10T^2}\big(X_3+2X_4\big)+\cO(F^5)~.
\ee


\section{Properties of the tensor $G^{abcd}$}\label{app:PropertiesofTensorG}

In this appendix we collect the key properties of the tensor field $G^{abcd}$, which contains second derivatives of the Lagrangian
\be
	G^{abcd}:=\big(L_{\a\a}F^{ab}+L_{\a\b}\~F^{ab}\big)F^{cd}+\big(L_{\a\b}F^{ab}+L_{\b\b}\~F^{ab}\big)\~F^{cd}~.
\ee
This tensor appears both in the equations of motion of NLED
\be
	L_{\a}(\pd_aF^{ab})+\hf G^{abcd}(\pd_aF_{cd})=0~,
\ee
and in the non-minimal operator relevant for quantisation
\be
	\D^{ab}_{\rm NLED}:=-L_{\a}\h^{ab}\Box+G^{acdb}\pd_c\pd_d+V^{acb}\pd_c~.
\ee
From its definition, $G^{abcd}$ inherits symmetries analogous to those of a curvature tensor:
\be
	G^{abcd}=-G^{bacd}~, \quad G^{abcd}=-G^{abdc}~, \quad G^{abcd}=G^{cdab}~.
\ee
However, the cyclic property only holds when the Lagrangian satisfies
\be
	G^{a[bcd]}=0 \quad \iff \quad \b(L_{\a\a}^2-L_{\b\b}^2)-2\a L_{\a\b}(L_{\a\a}+L_{\b\b})=0~,
\ee
as is the case for Maxwell and Born–Infeld electrodynamics. A further identity holds only in the case of conformal NLED \eqref{eqn:ConformalNLEDProperty}, when contracting two tensors in a single index
\be
	G^{abc}_{\ph{abc}d}\,G^{dfgh}=G^{abgh}\,G^{dcf}_{\ph{dcf}d} \quad \iff \quad L_{\a\a}L_{\b\b}-L_{\a\b}^2=0~.
\ee

In Section \ref{sct:QuarticHeatKernelTechniques} the tensor $G^{abcd}$ is realised as a background matrix $\dsG^{ab}$, its entries defined by
\be 
	(\dsG^{ab})^c_{\ph{c}d}:=G^{cab}_{\ph{cab}d}~.
\ee
Taking the trace gives
\be
	G^{ab}:=\tr \, \dsG^{ab}=2(\a L_{\b\b}-\b L_{\a\b})\h^{ab}+(L_{\a\a}+L_{\b\b})(F^2)^{ab}~.\label{eqn:trGForm}
\ee
For the case of conformal NLED \eqref{eqn:ConformalNLEDProperty}, $\dsG^{ab}$ inherits the following square property from its component tensor $G^{abcd}$
\be
	\dsG^{ab}\dsG^{cd}=G^{bc}\dsG^{ad}\quad \iff \quad L_{\a\a}L_{\b\b}-L_{\a\b}^2=0~.\label{eqn:dsG^2Prop}
\ee
This equation is key to solving momentum space integrals involving the exponential of the matrix $\cG(k):=-\ri k_ak_b\dsG^{ab}$
\baa
	\re^{\t\, \cG(k)}&=\id_4+\Bigg[\sum_{n=0}^{\infty}\frac{\t^{n+1}}{(n+1)!}\big(\tr\, \cG(k)\big)^n\Bigg]\cG(k)\non\\
	&=\id_4+\bigg[\frac{\re^{\t\, \tr\,\cG(k)}-1}{\tr\,\cG(k)}\bigg]\cG(k)~,
\eaa
where $\t$ is an arbitrary parameter.

From (\ref{eqn:trGForm}) it follows that the matrix $G:=(G^a_{\ph{a}b})$ is of composite form (\ref{eqn:4x4CompositeMatrix}). As shown in Appendix \ref{app:4x4MatrixProperties}, any such matrix satisfies the quadratic identity
\be \label{eqn:GQuadraticIdentity}
	G^2+2A G-C\id_4=0~,\qquad A:=-\tfrac{1}{4}\tr\,G~, \qquad C:=\tfrac{1}{4}\big[\tr(G^2)-\tfrac{1}{2}(\tr\,G)^2\big]~.
\ee
The quantities $A$ and $C$ can be expressed in terms of the Lagrangian derivatives as:
\baa
	A&=\a(L_{\a\a}-L_{\b\b})+2\b L_{\a\b}~,\\
	C&=4\a^2 L_{\a\a}L_{\b\b}+4\a\b L_{\a\b}(L_{\b\b}-L_{\a\a})+\b^2(L_{\a\a}+L_{\b\b})^2-4\b^2L_{\a\b}^2~.
\eaa

For causal NLED, the weak-field causality conditions (\ref{eqn:BackgroundConvexCondition}) lead to $C\geq0$. This follows directly from the condition $L_{\a\a}\geq0$:
\be
\left\{
\begin{aligned}
	&C=\b^2L_{\b\b}^2\geq0~,\hspace{13em} L_{\a\a}=0 \ \Rightarrow \ L_{\a\b}=0~,\\
	&\\[-1.5em]
	&C\geq \bigg(\frac{2\a L_{\a\a}L_{\a\b}+\b(L_{\a\b}^2-L_{\a\a}^2)}{L_{\a\a}}\bigg)^2\geq0~,\qquad L_{\a\a}>0 \ \Rightarrow \ L_{\b\b}\geq \frac{L_{\a\b}^2}{L_{\a\a}}~.
\end{aligned}
\right.\non
\ee
When restricting to causal NLED, to reflect the non-negativity of $C$ we adopt the notation
\be
	G^2+2A G-B^2\id_4=0~,\qquad B^2:=\sqrt{C}~.
\ee
Together, these relations characterise the algebraic role of $G^{abcd}$ in the NLED framework.


\newpage

\begin{footnotesize}

\end{footnotesize}

\end{document}